\documentclass[fleqn,usenatbib]{mnras}

\usepackage{newtxtext,newtxmath}
\usepackage[T1]{fontenc}

\DeclareRobustCommand{\VAN}[3]{#2}
\let\VANthebibliography\thebibliography
\def\thebibliography{\DeclareRobustCommand{\VAN}[3]{##3}\VANthebibliography}

\usepackage{graphicx}	% Including figure files
\usepackage{amsmath}	% Advanced maths commands

\usepackage{xcolor}
\usepackage{multirow}

\usepackage[dvipsnames]{xcolor}

\newcommand{\REV}[1]{{\color{black}#1}}

\newcommand{\REVcaption}[1]{%
  \caption{{\textcolor{black}{}}#1}
}

\newcommand{\aav}[1]{{\color{black}#1}}
\newcommand{\af}[1]{{\color{black}#1}}

\newcommand{\swone}{\texttt{S0.1}}
\newcommand{\swtwo}{\texttt{S40}}
\newcommand{\swthree}{\texttt{S200}}
\newcommand{\swfour}{\texttt{S40F}}

\graphicspath{ {./images/} }
\title{A Self-Consistent 3D Hydrodynamic Model for Helium Transit Signatures in Evaporating Hot Jupiters}
\author[A. Falorca and A. Vidotto]{
A. Falorca,$^{1}$\thanks{E-mail: falorca@strw.leidenuniv.nl }
A.~A.~Vidotto,$^{1}$
\\
$^{1}$Leiden Observatory, Leiden University, P.O. Box 9513, 2300 RA Leiden, The Netherlands\\
}

\date{Accepted XXX. Received YYY; in original form ZZZ}

\pubyear{\the\year{}}

\begin{document}
\label{firstpage}
\pagerange{\pageref{firstpage}--\pageref{lastpage}}
\maketitle

\begin{abstract}
The HeI triplet line (1083 nm), together with hydrodynamic models, can be used to characterize atmospheric escape of exoplanets. However, most of the available models  cannot capture the three dimensional (3D) physics of escaping atmospheres, such as tidal forces and the interaction with stellar winds. 
%what we did:
To investigate how 3D effects affect the helium transit signature, we update our 3D atmospheric evaporation model to self-consistently solve the hydrodynamic equations together with the atomic hydrogen and helium populations. 
We also produce synthetic helium transits.
Our atmospheric escape models assume a Hot Jupiter interacting with stellar wind of ranging mass-loss rates and two XUV fluxes, representative of an old and a young star.
%results:
Models considering an old star show a decrease of helium triplet density with increasing stellar wind strength, which occurs for two reasons.
First, stronger winds reduce the volume of the escaping atmosphere, which decreases obscuration of atmospheric transits.
Secondly, as a consequence of a less extended atmosphere, optical depth is reduced, impacting both photoionization  and heating, which in turn affect the gas temperature of planetary material, reducing the density of helium triplet.
%Since ionized helium recombination is the major source of triplet species, the evaporated material under higher temperatures leads to smaller column densities for the He I line.
The model assuming a younger star shows an extended outflow, with escape rates 25 times higher. 
For the same stellar wind strength, the helium transit is \REV{$3.3$} times deeper than when assuming the XUV of an older star. 
Weaker stellar winds and/or strong XUV flux allow for pre-transit helium absorption, while all scenarios show (different levels of) post-transit absorptions, described by the presence of a comet-like tail.
\end{abstract}


%%%%%%%%%%%%%%%%%%%%%%%%%%%%%%%%%%%%%%%%%%%%%%%%%%

%%%%%%%%%%%%%%%%% BODY OF PAPER %%%%%%%%%%%%%%%%%%

\section{Introduction}

% - introduce atmospheric escape
%
% - helium triplet as marker of atmospheric escape
%
% - observations
%
% - 3d models developed
%
% - stellar wind effects
%
% - structure of this paper

%*********************************************

Exoplanets orbiting close to their host stars are prone to undergo atmospheric escape.
This may occur through the process of hydrodynamic escape, likely to happen when the exoplanet is receiving a high flux of extreme ultraviolet radiation and X-rays (hereafter referred to as XUV radiation).
This is the driver of photoionization of gas such as hydrogen (H) and helium (He) present in the upper parts of the atmosphere.
The flux of stellar photons ionizes the planetary atmospheric gas, releasing their excess energy in the form of heat, causing the atmosphere to expand and overcome gravitational forces, resulting in a bulk outflow of atmospheric material \citep[][]{81_Watson, 03_Lammer}.

A planetary atmosphere that enters hydrodynamic (HD) escape will depend, among other factors, on the planetary gravity, the XUV flux received and the reservoir of gas material.
In extreme cases, the primordial atmosphere can be lost, drastically modifying the nature of the planet \citep[see e.g.][]{12_Lopez, 13_Owen, 17_Fossati, 18_Kubyshkina}.
It is the latter case that leads to theories that explain certain characteristics in population studies, such as the radius gap \citep{17_Fulton}, a scarcity of short orbital period exoplanets within a specific range of radii $1.5 - 2.0 R_\oplus$, or the Neptunian desert \citep{16_Mazeh}, where there is a lack of planets of short orbital period, within masses of $0.03-0.3 M_J$.
Additionally, planetary habitability can be significantly influenced by strong hydrodynamic escape \citep[][]{18_Lingam, 18_Dong}.

%i think the order of next paragraph can be swapped:
Transmission spectroscopy has been commonly used to detect hydrodynamic escape, with well studied hot Jupiters (HJ henceforth), such as HD209458b \citep{03_Vidal-Madjar} and HD189733b \citep[][]{10_Lecavelier, 12_Lecavelier, 12_Jensen, 13_BenJaffel} showing excess absorption in the hydrogen Lyman-$\alpha$ line, which probes all the way to the upper atmospheres of these planets. These exoplanets are ideal candidates for atmospheric studies due to their H- and He-rich atmospheres and large scale heights. In addition to hydrogen, a recent spectroscopic signature tracing escape, the metastable helium triplet (He I line) at $1083$ nm, has attracted considerable attention.
This spectral feature is produced by helium in the state $2^3\text{S}$ absorbing a photon, resulting in the transition of $2^3\text{S}$ to $2^3\text{P}$. 
For theoretical studies, see \citet{00_Seager} and \citet{18_Oklopcic}.
Since the first detections of escaping He \citep[see works of][]{18_Allart, 18_Nortmann, 18_Spake}, the number of detections has grown, including for several other exoplanets \citep[][]{19_Allart, 19_Alonso, 20_Guilluy, 20_Kirk, 21_Spake}.
This new marker has several advantages compared to the traditional hydrogen Lyman-$\alpha$ line: it is observed both in space-based facilities, as well as in ground-based ones.
In addition to this, the core of the He I line is not contaminated as is the case of the Lyman-$\alpha$ line, which suffers from absorption of interstellar medium material and contamination by geocoronal emission \citep{20_dosSantos}. %\citep[for a deepened discussion of the absorption by the ISM, see][]{09_Indriolo}.
%Encompassing this, transmission spectroscopy has been commonly used to detect hydrodynamic escape. %, as it causes excess absorption relative to the planet's obscuration.
%Several works have presented the community with such results for Hot Jupiters (HJ henceforth): HD209458b \citet{2003_Vidal} and HD189733b (\citet{10_Lecavelier, 12_Lecavelier}, \citet{12_Jensen}, \citet{13_BenJaffel}).
%These exoplanets are ideal candidates for atmospheric studies due to their H and He-rich atmospheres and large scale heights.

There have now been over 50 helium triplet transit observations, hence, population studies linking planetary parameters with the observed helium triplet absorption have become feasible \citep[e.g.,][]{24_Krishnamurthy, 24_Orell}.
However, there is now a large number of non-detections of helium transits in exoplanets that were expected to show such a signature. %several unresolved questions remain in this field of photoevaporation, primarily due to the high number of non-detections of atmospheric escape for exoplanets that possess characteristics expected to have a detectable escape.
For a compilation of detections and non-detections, we refer the reader to \citet{23_DosSantos}.
Possible physical phenomena that impact the predictions and detections of helium transits are the uncertainty in the stellar XUV fluxes rising from the challenge in observing and reconstructing XUV fluxes \citep{13_Linsky, 26_Behr} and also natural variability in stellar magnetic activity \citep{20_Guilluy, 23_Krolikowski, 25b_Allan}.
%; stellar He I variability has also gained traction \citep[][]{23_Krolikowski, 20_Fuhrmeister, 25_Mercier}; it is even valid to consider the use of inaccurate values of escape-impacting parameters belonging to the system at study.
%As an example, \citet{25_Mercier} studied the variability of solar He triplet: by exploring different causes, the authors saw that improper telluric correction could act as potential short-term source variability and stellar activity as definite long-term impact.
%The ability of stellar variability to mimic signatures of different physical nature is therefore another difficulty behind observations.
%Nevertheless, one thing that we want to investigate in the present paper is whether non-detections could be due to the presence of stellar wind.
Another physical process that could affect the helium transit signature (its potential detection and variability) is the presence of a stellar wind, which we investigate in our current study.

%Non-detections can also help setting constraints, for example, upper-limits in excess absorption of He I line \citep[][]{21_Krishnamurthy}, or by setting boundaries for $\text{He}/\text{H}$ fraction \citep[][]{22_Fossati}, or bounds to mass-loss rates \citep[][]{21_Zhang}.
%Despite observational difficulties, young age ($<1$ Gyr) exoplanets are expected to have the strongest escape \citep[][]{13_Owen}, hence the interest in this phase of planetary evolution.
%During this stage, high XUV flux and large planetary radius - which translates into a weaker gravitational potential - lead to strong atmospheric escape.

%Interestingly, the works of \citet{24_Krishnamurthy} and \citet{24_Orell} agree on the absence of a clear trend between atmospheric escape signatures and the system's age, contradicting predictions of several theoretical studies.
%As \citet{25_Allan} discussed in Section 3, for an individual planet along its evolutionary path, it should show stronger atmospheric escape and observational detectability at younger ages.
%Nonetheless, the observations that have been carried out so far have created a diverse sample of exoplanets such that the differences in the system's parameters impact strongly the triplet signature, as opposed to the contribution of only age.
%From these, both \citet{24_Krishnamurthy} and \citet{25_Allan} agreed on the importance of the geometric transit depth $(R_p/R_\star)^2$ related to detection of escape through the He I line.

Interactions with the wind of the host star shape the large-scale morphology of an escaping atmosphere \citep[for reviews on this topic, see e.g.][]{23_Kubyshkina, 25_Vidotto}.
From this interaction, several features can emerge: comet-like tails that trail behind the planet or streams of material headed towards the stellar surface, generally ahead of the planet's orbital motion.
The main characteristics of such structures depend on properties of the system, such as tidal forces, orbital velocity, and ram pressure of stellar wind \citep[][]{15_Matsakos, 17_CarrollNellenback}.
%By necessity, the effort to bring physical insight to features hiding in observations has increased: as an example, the work of \citet{25_Gascon} looks at the geometry of the trail of material around the exoplanet to extract more physical meaning from observations.
Additionally, atmospheric mass-loss rate of the exoplanet can also be affected by the stellar wind \citep[][]{20_Vidotto}.
To study the effects that stellar wind has on evaporating atmospheres, we must employ 3D numerical models \citep[e.g.][]{15_Matsakos, 17_CarrollNellenback, 21_Carolan}. 
However, there are still only a few 3D photoevaporation studies that include the He I marker 
\citep[][]{18_Wang, 20_Shaikhislamov, 21_a_Wang, 21_Wang, 22_MacLeod, 25_MacLeod, 24_Nail, 25_Nail}, some of which do not clearly address the influence of stellar winds in the He I line.

%This is the first 3D self-consistent radiative hydrodynamics model of photoevaporation focusing on H and He species in the same fluid, under the BATS-R-US framework.

In the current work, we perform 3D local hydrodynamic simulations of atmospheric escape in a close-in Hot Jupiter, including interaction with the stellar wind.
The novelty behind our work is the development of a self-consistent population-chemistry scheme that takes into account hydrogen and helium species.
We vary the strength of the stellar wind to investigate its effects on the transit spectral features of the He I line, as well as changing the stellar host.
%The paper is structured as follows: details of our 3D H-He radiation hydrodynamics model and parameters used in the simulations conducted in this study are described in Section \ref{section2}.
The paper is structured as follows: Section \ref{section2} details our \REV{3D H-He hydrodynamic model with incorporated radiation treatment,} while also presenting the parameters used in the simulations.
Our results are detailed in Section \ref{section4}.
We describe and compute synthetic observations of He I line transits in Section \ref{section5}.
A discussion is present in Section \ref{section6}, which includes future improvements, and finalize with conclusion in Section \ref{section7}.

\section{3D Modelling}
\label{section2}

% - main characteristics of 3d model
%
% - helium schematics
%
% - photoionization steps
%
% - boundaries and grid

%*********************************************

We model atmospheric escape of a Hot Jupiter (HJ) by building upon the works of \citet{21_Carolan} and \citet{23_Allan}.
The model presented here is an extensive update of \citet{21_Carolan}, where we now include He species in the previous H-only three-dimensional (3D) HD models.
The helium physics follows a close implementation of \citet{23_Allan} and is self-consistently computed (i.e., coupled) with the HD equations.

\subsection{Fluid dynamics}

We use BATS-R-US \citep{05_Toth} to simulate  hydrodynamic escape via photoevaporation of the atmospheres of close-in exoplanets, including the space environment around them. This code has assisted in the study of stellar winds interacting with planetary outflow in several previous works of our group \citep[e.g.,][]{21_a_Carolan, 21_Carolan, 21_Hazra, 22_Kubyshkina, 24_Presa}.
The present model uses a multi-species fluid, totaling to 10 the number of hydrodynamic variables in the model:
\begin{itemize}
    \item Species mass densities: $\rho_\text{sp}$, where $\text{sp}$ can stand for: neutral hydrogen $\text{H}_0$, ionized hydrogen $\text{H}^+$, helium singlet $\text{He}(1^1S)$, helium triplet $\text{He}(2^3S)$, ionized helium $\text{He}^+$ and fully ionized helium $\text{He}^{++}$.
    \item Velocity $\vec{v} = $($v_x$, $v_y$, $v_z$).
    \item Thermal pressure ($P_\text{th}$).
\end{itemize}
We iteratively solve the set of hydrodynamic equations of mass conservation
\begin{equation}
    \frac{ \partial \rho }{\partial t} + \nabla \cdot \left( \rho \vec{v} \right) = 0 \,
    \label{EQ--mass-conservation}
\end{equation}
momentum conservation
\begin{equation}
\begin{split}
    \frac{ \partial \left( \rho \vec{v} \right) }{\partial t} + & \nabla \cdot \left( \rho \vec{v}\vec{v} + P_\text{th} \hat{I} \right) \\ 
    & = \rho \left( \vec{g} - \frac{G M_\star}{\left(r-a \right)^2 } \hat{R} - \vec{\Omega}\times \left( \vec{\Omega}\times \vec{R} \right) - 2\left(\vec{\Omega}\times\vec{v} \right) \right)\, ,
\end{split}
\label{EQ--momentum-conservation}
\end{equation}
and energy conservation
\begin{equation}
\begin{split}
    \frac{ \partial \epsilon }{\partial t} + & \nabla \cdot \left( \vec{v} \left( \epsilon + P_\text{th} \right) \right) \\
    & = \rho \left( \vec{g} - \frac{G M_\star}{\left(r-a \right)^2 } \hat{R} - \vec{\Omega}\times \left( \vec{\Omega}\times \vec{R} \right) \right) \cdot \vec{v} + \mathcal{H} - \mathcal{C} \, ,
\end{split}
\label{EQ--energy-conservation}
\end{equation}
where $\rho$ is the total mass density ($\rho = \sum_\text{sp} \rho_\text{sp}$), $\hat{I}$ the identity matrix, $\vec{r}$ and $\vec{R}$ the position vectors relative to the center of the planet or the star, respectively, $\Omega$ the orbital rotation rate, $M_\star$  the stellar mass and $\vec{g}$ is the acceleration due to the plane's gravity.
Throughout this work, constants related to the stellar host have a subscript `$\star$' and the HJ have subscript `p', unless further specification is given. The total energy density $\epsilon$ is given by:
\begin{equation}
    \epsilon = \frac{1}{2} \rho v^2 + \frac{P_\text{th}}{\gamma-1} \, ,
    \label{EQ--energy}
\end{equation}
where $\gamma$ is the adiabatic index, set to $5/3$.
The energy conservation (Equation (\ref{EQ--energy-conservation})) contains the terms of volumetric heating rate ($\mathcal{H}$) and volumetric cooling rate ($\mathcal{C}$), which will be discussed in Subsection \ref{subsection_Photoionization and Heating}.

% And since the model is using multi-species in the same fluid:
% \begin{equation}
%     \rho = \sum_\text{sp} \rho_\text{sp}
%     \label{EQ--density}
% \end{equation}
% where the subscript '$\text{sp}$' stands for individual species.

\subsection{Hydrogen and Helium state populations}

In addition to the previous equations, the model also solves six continuity equations for the population of helium and hydrogen
\begin{equation}
    \frac{ \partial \rho_\text{sp} }{\partial t} + \nabla \cdot \left( \rho_\text{sp} \vec{v} \right) = \left[ \mathcal{S} - \mathfrak{S} \right]_\text{sp} \, ,
    \label{EQ--population}
\end{equation}
where $\mathcal{S}_\text{sp}$ and $\mathfrak{S}_\text{sp}$ represent the sources and sink rates for each species, respectively. Notice that if we sum over all species the Equation (\ref{EQ--population}), we obtain Equation (\ref{EQ--mass-conservation}).

\begin{figure*}
	\includegraphics[width=0.75\linewidth]{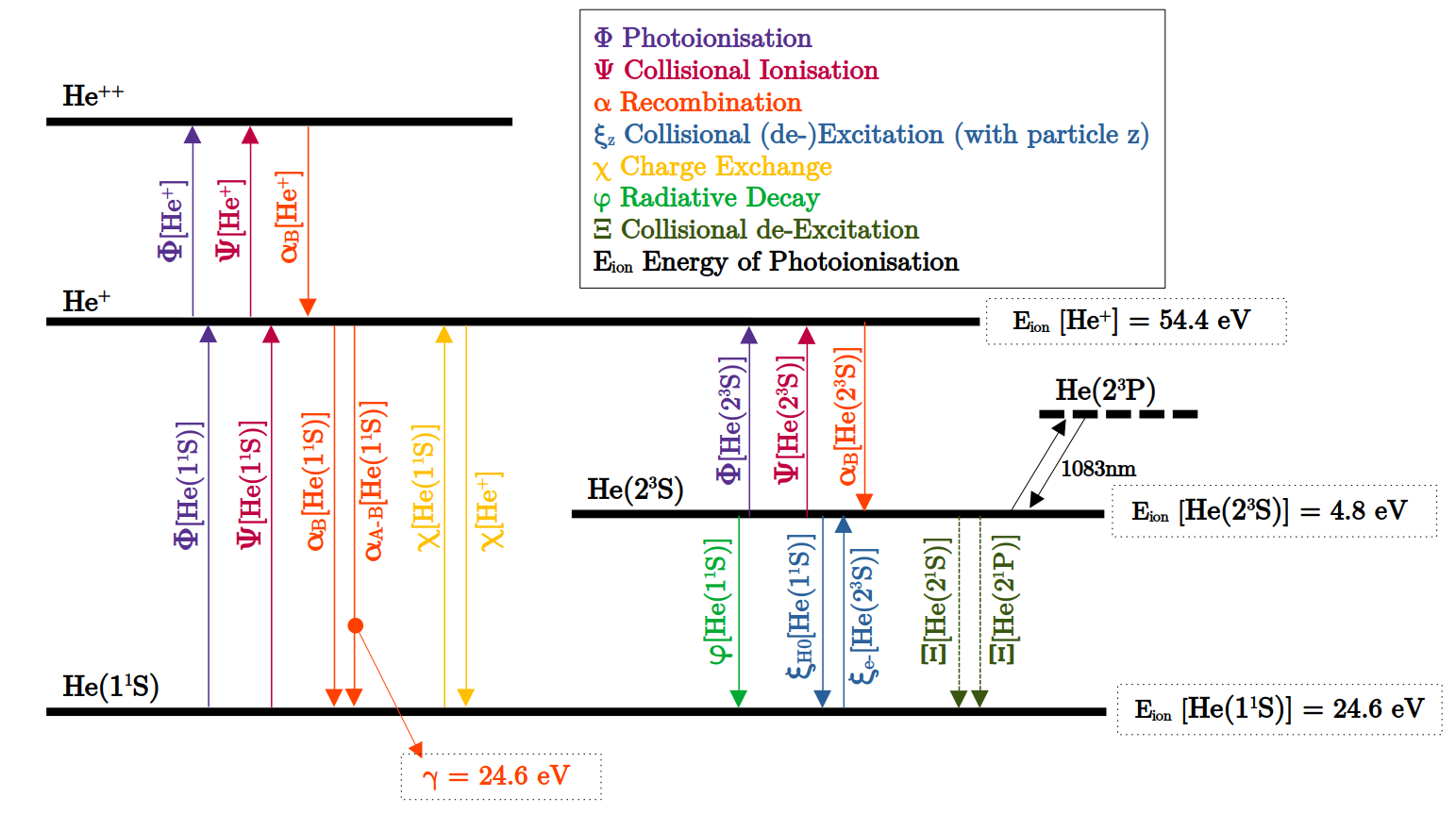}
    \caption{Schematic representation of production or annihilation of different He states.
    }
    \label{fig:He1-schematic}
\end{figure*}

The reaction rates are described in Table \ref{tab:rates} and a schematic representation for the helium population is shown in Figure \ref{fig:He1-schematic}.
We consider the following reaction rates: photoionization ($\Phi$); collisional ionization ($\Psi$); recombination ($\alpha$); collisional excitation and collisional de-excitation ($\xi$ and $\Xi$); radiative decay ($\varphi$) and charge exhange ($\chi$).
The sources and sinks of each species (Equation (\ref{EQ--population})) are calculated as follows:
\begin{itemize}
    \item The hydrogen population (i.e., species $\text{H}_0$ and $\text{H}^+$) has
    \begin{align}
    \left[ \mathcal{S} - \mathfrak{S} \right]_{\text{H}_0} & = + \Big{ \{ }
    \alpha_B \, [\text{H}_0] + \text{A}_\text{H} \chi \, [\text{He}^+]  \Big{ \} } \notag \\ 
    & - \Big{ \{ } \Phi[\text{H}_0] + \Psi \, [\text{H}_0] + \text{A}_\text{H} \chi \, [\text{He}(1^1S)] \Big{ \} }
    \label{EQ--population1} \, ,
    \end{align}
    and
        \begin{align}
    \left[ \mathcal{S} - \mathfrak{S} \right]_{\text{H}_0} = -\left[ \mathcal{S} - \mathfrak{S} \right]_{\text{H}^+} \, .
    \end{align}

    \item For the helium singlet ground state:
    \begin{align}
    & \left[ \mathcal{S} - \mathfrak{S} \right]_{\text{He}(1^1S)}  = + \Big{ \{ }
    \alpha_B \, [\text{He}(1^1S)] + \alpha_{A-B} \, [\text{He}(1^1S)] \notag \\
    & + \Xi \, [\text{He}(1^1S)]  + \text{A}_\text{He} \chi \, [\text{He}(1^1S)] + \xi \, [\text{He}(2^1S)] + \xi \, [\text{He}(2^1P)]  \notag \\
    & + \varphi \, [\text{He}(1^1S)] \Big{ \} } - \Big{ \{ } \text{A}_\text{He} \chi \, [\text{He}^+] + \Phi \, [\text{He}(1^1S)] + \Psi \, [\text{He}(1^1S)]  \notag \\
    & + \Xi \, [\text{He}(2^3S)] \Big{ \} }  \, .
    \label{EQ--population2}
    \end{align}

    \item For the helium triplet state:
    \begin{align}
    & \left[ \mathcal{S} - \mathfrak{S} \right]_{\text{He}(2^3S)} =  +\Big{ \{ } \alpha_B \, [\text{He}(2^3S)] + \Xi \, [\text{He}(2^3S)]  \Big{ \} } \notag \\
    & - \Big{ \{ } \xi \, [\text{He}(2^1S)] + \xi [\text{He}(2^1P)] +  \Xi \, [\text{He}(1^1S)] + \varphi \, [\text{He}(1^1S)] \notag \\
    & + \Phi \, [\text{He}(2^3S)] + \Psi \, [\text{He}(2^3S)] \Big{ \} }  \, .
    \label{EQ--population3}
    \end{align}

    \item For the singly ionized helium state:
    \begin{align}
    & \left[ \mathcal{S} - \mathfrak{S} \right]_{\text{He}^+}  = + \Big{ \{ } \Phi \, [\text{He}(1^1S)] + \Phi \, [\text{He}(2^3S)]  \notag \\
    & +  \Psi \, [\text{He}(1^1S)] + \Psi \, [\text{He}(2^3S)] + \alpha_B \, [\text{He}^+] + \text{A}_\text{He} \chi \, [\text{He}^+]  \Big{ \} }  \notag \\ 
    & - \Big{ \{ } \alpha_B \, [\text{He}(1^1S)] + \alpha_{A-B} \, [\text{He}(1^1S)] + \alpha_B  \, [\text{He}(2^3S)]  \notag \\
    & +  \text{A}_\text{He} \chi \, [\text{He}(1^1S)] + \Phi \, [\text{He}^+] + \Psi \, [\text{He}^+] \Big{ \} }  \, .
    \label{EQ--population4}
    \end{align}
    
    \item For the doubly ionized helium:
    %\begin{equation}
    \begin{align}
    \left[ \mathcal{S} - \mathfrak{S} \right]_{\text{He}^{++}} = + \Big{ \{ } 
     \Phi \, [\text{He}^+] + \Psi \, [\text{He}^+] \Big{ \} } - \Big{ \{ }  
     \alpha_B \, [\text{He}^+]  \Big{ \} }   \, .
    \label{EQ--population5}
    \end{align}
    %\end{equation}

There is a detail on the reaction of charge exchange $\mathbf{\chi}$, where hydrogen and helium react together and form different final species.
Importantly, in order to preserve mass continuity, we distinguish charge exchange between helium species and hydrogen species by taking the correct species' mass number ($\text{A}_\text{H} = 1$, $\text{A}_\text{He} = 4$).
\af{
The use of mass number allows us to maintain units of volumetric rate times mass $(\text{g}\,\text{cm}^{-3}\,\text{s}^{-1})$ for the entirety of the scheme, which justifies the presence of $m_\text{H}$ in the charge exchange rates $\chi[\text{sp}]$ present in Table \ref{tab:rates}.
}
%By writing Equation (\ref{EQ--population}) in terms of number density, consequently transforms rates present in Table \ref{tab:rates} ($\chi$) into number volumetric rates (lets use $\bar{\chi}= \chi / (m[\text{sp}])$, where $m[\text{sp}]$ stands for the mass of the species).
%Thus, when the focus is on depopulating ionized He, we use $\chi [\text{He}(1^1S)] = \bar{\chi} [\text{He}(1^1S)] \times m[\text{He}] = 1.25 \times 10^{-15} \left(\frac{T}{300} \right)^{0.25} \text{n}_{\text{H}_0} n_{\text{He}^+} \times m[\text{He}]$, and similarly, when the focus is on depopulating neutral H, we use $\chi [\text{He}(1^1S)] = \bar{\chi} [\text{He}(1^1S)] \times m[\text{H}]$.

% Additionally, many of the reaction rates present in Table \ref{tab:rates} should have temperature distinction between electron, ion or fluid temperatures. 
% Due to the nature of the model, we can only retrieve a fluid averaged temperature.

\af{
Although we closely follow the calculation of the helium population of \citet{23_Allan}, we do not account for the population of $2^1S$ and $2^1P$ states \citep[similar to other studies in the literature:][]{18_Oklopcic, Lampon--2020, 22_DosSantos}.
}
%The radiative decay rate of $\text{He}(2^1P)$ follows $\varphi [\text{He}(2^1P)] = 1.8\times10^9 n_{\text{He}(2^1P)} \text{ cm}^{-3}\text{s}^{-1}$ \citep[][]{09_Wiese}, and $\varphi [\text{He}(2^1S)] = 50.94 n_{\text{He}(2^1S)} \text{ cm}^{-3}\text{s}^{-1}$ \citep[present in CHIANTI database:][]{25_Schulik}.
Our assumption of immediate transition from the $\text{He}(2^1S), \text{He}(2^1P)$ states to the ground one $\text{He}(1^1S)$ is justified by the fact that the radiative decay rate of the metastable state is several orders of magnitude smaller.
%These are several orders of magnitude higher than the radiative decay of the metastable state, thus, our assumption of immediate transition from the  $\text{He}(2^1S), \text{He}(2^1P)$ states to the ground one $\text{He}(1^1S)$.
\af{
In this context, however, 1D works that have included these the $\text{He}(2^1S)$ and $\text{He}(2^1P)$ states \citep[][]{23_Allan, 25_Munoz, 25_Schulik}, did not find any significant impact on the triplet population, as these states instantly become depopulated.
%Nevertheless, these intermediate states tend to recombine or radioactively decay, which can also emit photons capable of ionizing the helium triplet.
}
%The quantum nature underlying recombination rates requires the distinction between Case-A (direct) and Case-B (indirect) coefficients.
%As a higher energy state recombines to a possible combination of lower energy states until reaching the ground state, the intermediate processes can emit photons - some of which, can potentially ionize hydrogen and helium states.
%The direct case is therefore the summation over all possible transitions, while $\alpha_\text{B}$ is the summation over all possible transitions without the direct transition to the ground state.
%Therefore, the adopted notation of $\alpha_\text{A-B}$ describes the case of recombination directly to the ground state.
%The importance of this reaction is raised when the fluid is optically thin or thick: for an optically thick fluid, it is typical to factor in $\alpha_\text{B}$, as the released photons ionize locally the same state that has recombined, impacting the population to a near net change.

\end{itemize}

\subsection{Photoionization and Heating/Cooling Processes} \label{subsection_Photoionization and Heating}

Compared to the monochromatic model of \citet{21_Carolan}, we now separate the incoming radiation into four energy bins, similar to \citet{21_a_Wang,23_Allan}. The resulting wavelength ranges are X-rays, hard EUV (hEUV), soft EUV (sEUV) and mid UV (mUV).
We distinguish these wavelength channels by associating each one with a photon energy $\mathfrak{e}_\lambda$: $\mathfrak{e}_\text{mUV} = 7.0 \text{ eV}$; $\mathfrak{e}_\text{sEUV} = 20.0 \text{ eV}$; $\mathfrak{e}_\text{hEUV} = 40.0 \text{ eV}$; $\mathfrak{e}_\text{Xray} = 248.0 \text{ eV}$.
%These photoionization processes will help determine the population of species and the amount of heat inserted into the atmosphere.
The literature values of photoionization energy cost for each species $\text{E}_\text{sp}$ are: $\text{E}_{\text{H}_0} = 13.6 \text{ eV}$; $\text{E}_{\text{He}(1^1S)} = 24.6 \text{ eV}$; $\text{E}_{\text{He}(2^3S)} = 4.8 \text{ eV}$ and $\text{E}_{\text{He}^+} = 54.4 \text{ eV}$. 

The model requires values of individual optical depths $\tau_{\lambda}$ to account for photoionization rates and heating.
This is employed under the plane-parallel ray approximation.
\begin{equation}
    \tau_{\lambda} = \sum _{\text{sp}} \tau_{\lambda}[\text{sp}] = \sum _{\text{sp}} \int _{\text{start } x} ^{\text{final } x} n_\text{sp} \sigma_{\text{sp}, \lambda} dx \, ,
    \label{EQ--optical-depth}
\end{equation}
where $\sigma_{\text{sp}, \lambda}$ is the wavelength-dependent photoionization cross-section, valid only when $\text{E}_\text{sp} \leq \mathfrak{e}_{\lambda}$, i.e., the species is able to be ionized by a photon.
%It is calculated using theoretical fits as follows: $\text{H}_0$, $\text{He}(1^1S)$, $\text{He}^+$ are based on the fits provided by \citet{Verner--1996}, detailed in Table \ref{tab:verner_params}.
%The cross-section of $\text{He}(2^3S)$ is present in Table \ref{tab:BIassoni_params}, using parameters outlined in \citet{Biassoni--2023}.
%Notice additionally that the integration is performed starting at the leftmost boundary ($-x$) from where the flux is coming.
\af{
The integration is performed starting at the leftmost boundary ($-x$) from where the flux is coming.
The cross-sections for species $\text{H}_0$, $\text{He}(1^1S)$, $\text{He}^+$ follow the theoretical fits provided by \citet{Verner--1996}.
The triplet species $\text{He}(2^3S)$ uses the same parameters outlined in \citet{Biassoni--2023}.
}
\REV{The plane-parallel approximation performs computationally well in our Cartesian grid, however it can have non-physical effects on the extension of the  photoionization and heating deposition.
For example, the nightside of the planet should see an umbra cone extended up to $\sim a/(\frac{R_\star}{R_p}-1) \simeq 10 R_p$, instead of an infinite cylinder of no radiation.
Given that the density of material falls off very quickly, at $\sim 10 R_p$ in the nightside of the planet, the material there has such low density that its contribution should be negligible to the synthetic transits we later compute.
Additionally, there is a loss of flux conservation, since the plane-parallel radiation does not follow the inverse square law.
This only corresponds to a difference of $\lesssim 20\%$ at distances of the order of $R_\star$, impacting minimally heating and photoionization closer to the planet. 
The work of \citet{22_Yan} has compared in more detail spherical-spherical radiative transfer model to a plane-parallel model, finding that the latter can lead to enhanced absorption features. %: i.e., the mass volumetric rates and heating volumetric rates would not see a change in their overall relative rate hierarchy.
}

The rate of photoionization for a given species and for a given wavelength-channel is given by:
\begin{equation}
    \Phi_{\lambda}[\text{sp}] = \mathcal{F}_{\lambda} \omega_{\text{sp}, \lambda} \exp{(-\tau_{\lambda})} \sigma_{\text{sp}, \lambda} \frac{1}{\mathfrak{e}_{\lambda} }  \, ,
    \label{EQ--photoionization}
\end{equation}
where $\mathcal{F}_{\lambda}$ is the flux of the corresponding wavelength and $\omega_{\text{sp}, \lambda}$ is the weighting factor that prevents the same photon from photoionizing multiple species, as defined in \citet{23_Allan}:
\begin{equation}
    \omega_{\text{sp}, \lambda} = \frac{ n_\text{sp} \sigma_{\text{sp}, \lambda} }{ \sum_{\text{sp}} n_\text{sp} \sigma_{\text{sp}, \lambda} } \, .
    \label{EQ--weighting}
\end{equation}
If we sum $\Phi_{\lambda}[\text{sp}]$ over the photoionization channels, we obtain $\Phi[\text{sp}]$, as present in Equations (\ref{EQ--population1}) to (\ref{EQ--population5}).
We use a specific numerical treatment to calculate $\exp{\left(-\tau_\lambda\right)}$, as a means to avoid truncation errors, described in Appendix \ref{appendix_exp-tau}.

The volumetric heating rate that is inserted in the atmosphere by the excess energy from the photoionizing photons is given by:
%\begin{equation}
\begin{align*}
\mathcal{H}_{\text{sp}, \lambda} & = \delta_{\text{sp}, \lambda} \mathfrak{e}_{\lambda} \Phi_{\lambda}[\text{sp}] n_{\text{sp}}  \\ & = \delta_{\text{sp}, \lambda} \mathcal{F}_{\lambda} \omega_{\text{sp}, \lambda} \exp{(-\tau_{\lambda})} \sigma_{\text{sp}, \lambda} n_{\text{sp}}   \, ,  
\end{align*}
\label{EQ--heating}
%\end{equation}
where $\delta_{\text{sp}, \lambda} = \left( 1 - \text{E}_\text{sp} / \mathfrak{e}_{\lambda} \right)$ accounts for the proportional excess in energy.
The heating term incorporated in Equation (\ref{EQ--energy-conservation}) corresponds to the summation of $\mathcal{H}_{\text{sp}, \lambda}$ over all species and wavelengths.
\REV{Here, each channel's flux value $\mathcal{F}_{\lambda}$ is constant over the entire grid: it is the optical depth term $\exp(-\tau)$ that dynamically informs the model about the local absorption of the incoming radiation.
}

\af{
In addition to the stellar flux, our model also accounts for the contribution of photons originated in the recombination of $\text{He}^+$ to $\text{He}(1^1S)$, with $\mathfrak{e}_{\lambda[\alpha]}=24.6\, \text{eV}$, as done in \citet{23_Allan}.
%So, we assume a photoionization rate of the same value - we are essentially expressing that locally, these photons are immediately re-absorbed.
%Since these photons will be predominantly released in an optically thick region, close to the planet's surface, we take this assumption as reasonable.
%These photons will impact the population scheme, bringing into addition the amount of weighted photoionizations $\xi_{\text{sp},\text{ch}} \alpha_{\text{A-B}}$ to the species possible to ionize, and additional heating.
%We assume that these released photons act locally in the photoionization and heating processes.
We assume that these released photons act locally, produced with a rate equal to the weighted rate of the recombination process $ \Phi_{\lambda[\alpha]} [\text{sp}] = \omega_{\text{sp}, \lambda[\alpha]}  \alpha_\text{A-B}$.
This contribution is added to the photoionization rate of the affected species ($\text{H}_0, \, \text{He}(1^1S), \, \text{He}(2^3S)$) and the same applies to the excess heating produced.
%and follow the photon weighting for the respective recombination rate. %, in addition to the stellar flux that is proceeded in a parallel-ray approximation.
}

Moreover, our model computes 12 volumetric cooling rates shown in Table \ref{tab:cooling_processes}.
These contributions come from hydrogen and from helium collisions - both collisional ionization and collisional excitation - and recombination.
Free-free emission cooling is stated in a way that already entails the entire species population contribution.
These cooling terms were derived by \citet{Black--1981} and \citet{Cen--1992}, but we use the same transformation present in \citet{23_Allan} for the collisional excitation of helium triplet.
\REV{In essence, this modification uses the explicit number density of $\text{He}(2^3S)$ that the code computes, instead of assuming a steady-state solution dominated by recombination ($ n_{\text{He}(2^3S)}\propto T^{-0.6687} n_\text{e} n_{\text{He}^+}$) as done by \citet{Black--1981} - this would have been incorrect in our case.
}
It is the summation of all these cooling terms that make up the cooling term $\mathcal{C}$ in Equation (\ref{EQ--energy-conservation}).

\af{
As a final remark, reaction rates and volumetric heating and cooling rates are not computed for fluid predominantly ($\gtrsim 98\%$) composed of stellar wind.
This condition is defined using a passive scalar variable, which distinguishes gas that has advected from a planetary or stellar boundary condition.
}

\subsection{Simulation Setup}
%\subsection{Boundaries and Grid}

In our models, we use a cartesian grid with the following limits: $x, y = [-48, 48] \, R_p$ and $z=[-32, 32] \, R_p$.
The planet is located at the origin of this grid, defined by fixed inner boundary conditions at $r=1.0\, R_p$.
In this inner boundary, we set fixed base number density $n_0$ and temperature $T_p$, and the velocity uses a reflective  boundary condition.
We assume the Hot Jupiter to be tidally locked, such that the star is always in the $-x$ direction, while the planet orbits in the $+y$ direction, on the plane $z=0$.
At the outer boundaries, all variables are under inflow limiting boundary condition, except for the $yz$ plane at the $-x$ boundary.
\REV{In the latter, we follow the procedure from \citet{21_a_Carolan}: we inject the stellar wind assuming a 1D isothermal model \citep{65_Parker}, which is computed by choosing two free parameters, the stellar wind mass-loss rate ($\dot{\mathcal{M}}_\star$) and stellar wind temperature ($T_\star$), which in turn set both the density and velocity radial profiles.
Having the stellar outflow profile, we orient the stellar wind at the $-x$ boundary to flow radially away from the star, also taking into account the Coriolis force $-2 {\vec{\Omega}} \times{\vec{v}} $ and centrifugal force $-\vec{\Omega} \times ( \vec{\Omega} \times \vec{r} )$, necessary to convert into the planet's reference frame. 
}

The HD equations are solved using a second-order numerical scheme with Linde flux limiter and minmod slope limiter to calculate the Riemann Problem fluxes at cell interfaces.
Advancing the time step follows an explicit second-order temporal scheme with a Courant-Friedrichs-Lewy (CFL) number set to 0.25.

\REV{All our models are initially ran in ‘Steady-State Mode’, i.e., different time steps are used in different grid cells – limited by the local stability conditions.
This method accelerates convergence towards steady state, but it does not have a physical temporal information.
We typically run our models for at least $60$ thousand iterations until simulations reach global steady state.
After that, we run our models for an additional $2$-minute of physical evolution (i.e., using ‘Time Accurate Mode’), so that the metastable population converges into a smooth solution.
}

\REV{There are a couple of procedures we implement in our runs to verify and mitigate potential spurious effects in the outflow dynamics. 
For example, we start our models only incorporating the planetary outflow (including XUV irradiation, but no orbital motion, no rotation). 
We run the code until this is approximately in steady state (around 6 to 8 thousand iterations).
We compare this output with outputs of 1D models to confirm they have the same solution. 
Then, we turn on Coriolis and tidal forces and again wait until the solution reaches steady state (another 6 to 8 thousand iterations). 
Only after this we inject the stellar wind in the grid. 
This is the usual approach in our models. 
This has two benefits: 1) we carefully check the evolution of the dynamics at each intermediate setup; and 2) we save computational resources as we can use the same output at iteration $\sim 16000$ for multiple runs with different stellar wind conditions. 
In some cases, we can further speed convergence by using the final output of one model (\texttt{S8.0}) as input to another model (\swone).
}

\begin{figure}
    \centering
        \includegraphics[width=0.8\linewidth]{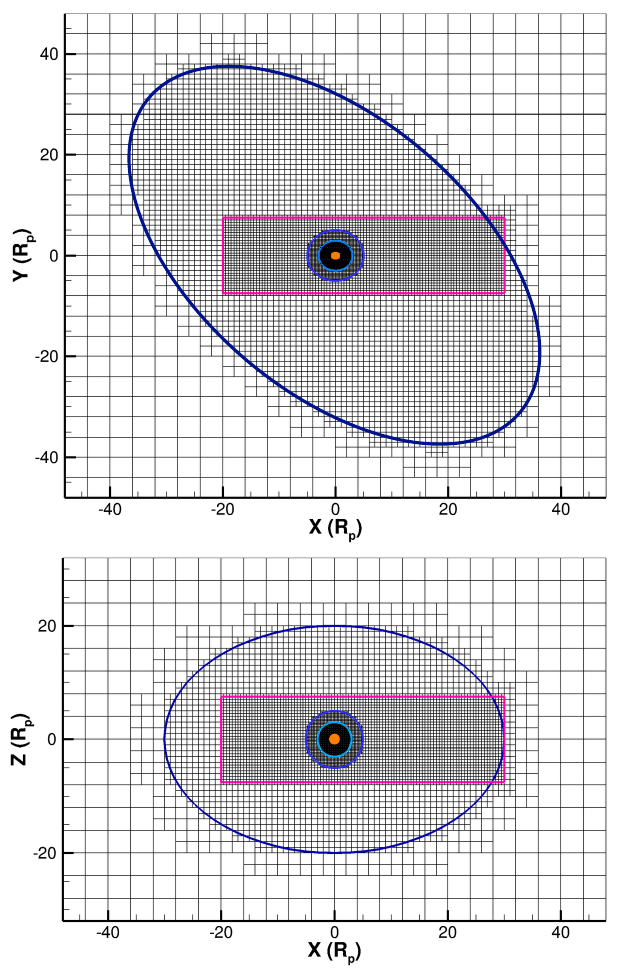}
    \caption{Grid blocks in $z=0$ (top) and $y=0$ (bottom) planes showing the different resolution regions assumed in our simulations. 
    The orange circle depicts the Hot Jupiter boundaries; the light blue circle defines the highest resolved part of the grid, the sphere with radius $3\, R_p$, followed by a blue circle showing the next resolved concentric sphere of radius $5\, R_p$; then is the horizontal cylinder marked by the pink line, and finally, the rotated elongated ellipsoid as the last resolved part of the grid, the large dark blue ellipsoid.
    Each grid block has $4\times4\times4$ cells.}
    \label{fig:grid}
\end{figure}

%The grid is initialized using adaptive mesh refinement, evolving its local resolution in the first few iterations to a desired configuration.
\REV{Previous to any numerical step, the grid starts in its minimal resolution, defined by the extent of the Cartesian domain and the number of blocks of cells for each direction. 
Then, we let the adaptive mesh refinement evolve the base grid until reaching the desired (maximum) level of refinement in each region.
%When these 10 iterations are finished, we keep the newly resolved grid fixed for the rest of the run.
The adopted regions are:
}
\REV{\begin{itemize}
    \item Around the planet, we have a resolution of $1/32 \, R_p$ in a shell between $r=R_p$ and $r=3\, R_p$.
    \item Adjacently, follows another shell with cell resolution $1/16\, R_p$, until $r=5\, R_p$.
    \item These shells are inscribed in a cylinder, with its central axis pointing in the $x$ direction, with a radius of $7.5\, R_p$ and a height of $50\, R_p$, centered at $\vec{r}=(5, 0, 0)\, R_p$. 
    This cylinder possesses a cell resolution of $1/8 \, R_p$.
    \item We implemented an additional rotated ellipsoidal region of resolution $1/4\, R_p$. 
    The latter geometry allows us to better resolve the planetary outflow under the influence of orbital motions and tidal forces.
\end{itemize}
Once these refinements are finished, we turn off the adaptive mesh refinement and the grid remains the same until the end of the run. 
At each level of refinement, a certain cell is split into 2 in each direction  (therefore increasing the resolution by a factor of 2, with the old cell splitting into 8 cells).
Only one level of refinement is added (or removed) at an iteration, which implies we perform a series of refinements. 
We experimented with how frequently the subsequent refinement levels can be implemented without breaking the code. 
Usually, the numerical computation runs smoothly when we build the grid within the first 10 iterations (in this case, there is nearly no evolution on the initial conditions set on the grid). 
However, past tests also showed that, depending on the characteristics of the system and the assumed initial conditions, a slower build of the grid might be necessary.% (e.g., increasing one resolution level at a time at every 10 - 100 iterations), but for our models, this ``slow build'' was not necessary. 
}
Figure \ref{fig:grid} displays the different levels of resolution in the grid.
In total, each simulation has around $19.7$ million cells.
We point out that the resolution applied in these models is much higher than the previous models of pure H atmosphere \citep[][]{21_a_Carolan}, as the study of complex He-physics scheme required higher resolution.

%Final ideas on the model.

Altogether, Equations (\ref{EQ--mass-conservation}) to (\ref{EQ--energy-conservation}) describe the fluid dynamics, coupled to the population balance Equations (\ref{EQ--population}) to (\ref{EQ--population5}).
In summary, our model considers photoionization, recombination, collisional ionization, excitation and de-excitation, charge exchange and radiative decay - as defined in Table \ref{tab:rates}.
In addition to this, cooling and heating are also calculated simultaneously with the scheme's species population.

\subsection{Modelling Parameters}

%\section{Simulations}
\label{section3}

% - Inputs - describe how i do it
% 
% - Family of models
%
% - 3D output
%
% - 2D output
%

%*********************************************

%\subsection{Parameters for Setup}

The photoionization process is the driver of the escaping atmosphere, hence the importance of XUV flux from the host star.
The incorporated fluxes for the two solar-like stars considered in this study are detailed in Table \ref{tab:models_props} as input parameters.
The first star (Star 1) has a mass of $M_{\star,1} = 1.1 M_\odot$, radius of $R_{\star,1} = 1.2 R_\odot$ and a stellar spectrum taken from \citet{16_France} (see Figure 3 of \citet{23_Allan}), corresponding to an old ($\sim5.6$ Gyr) K-star.
%The spectrum is taken from the K6 star, HD 85512, which is estimated to be $~5.6\pm0.1$ Gyr with a luminosity of $\log_{10}(L_\text{Bol})=32.78$, less than an order of magnitude smaller than our solar luminosity.
The second star (Star 2) has a larger mass $M_{\star,2} = 1.3 M_\odot$ but smaller radius $R_{\star,2} = 1.1 R_\odot$ \citep[][]{08_Vauclair}.
The spectrum is presented in \citet{25b_Allan}, corresponding to $\iota$-Hor during a maximum phase of its activity cycle. 
Because this star is young ($\sim625$ Myr), the fluxes in all channels are much higher than the first star, in particular the mid-UV and X-ray channels.

%In addition to the stellar flux, our model requires relative abundances as inputs.
%There is a large number of free parameters in our models, the main cause being the existence of two distinctly originated fluids: the atmospheric material from the HJ hydrodynamic escape, and the stellar wind.
To explore the differences in the adopted stellar spectra, our simulated exoplanet is unaltered: it orbits the host star at $a=0.05\text{AU}$, has a mass of $M_p=0.67M_J$ and a radius of $R_p=1.37R_J$.
The Hill radius present in Table \ref{tab:models_props}, $r_\text{H} = a \left(M_p/3M_\star\right)^{1/3} \sim 6 \, R_p$ is similar in both stellar inputs.
At the base of the (upper) atmosphere, we consider a fixed temperature $T_p=1000\text{K}$ and a base density $\rho_0 = 3 \times10^{-13}\, \text{g}\, \text{cm}^{-3}$. 
These values are informed by the values at the top of the lower atmosphere from 1D models of \citet{23_Allan}.
In addition, we fix the species ratios in our inner boundary.
The ratio of He to H in the base of the planetary outflow is $[\text{He}/\text{H}]_p=\left( n_\text{He} / n_\text{H} \right)_p = \frac{1}{9}=0.111$, a ratio higher than the one applied to the stellar wind $[\text{He}/\text{H}]_\star=0.068$, resembling our solar values \citep[][]{07_Kasper}.
The individual species ratios are written with respect to $\text{n}_{\text{H}}$, and ordered as $\{ \text{H}_0 : \text{H}^+ : \text{He}(1^1S) : \text{He}(2^3S) : \text{He}^+ : \text{He}^{++} \}$:
\begin{itemize}
    \item At the planetary launch base: $ \{ \sim (1.0-[\text{He}/\text{H}]_p) $ : $ 10^{-5} $ : $ 0.9[\text{He}/\text{H}]_p $ : $ 10^{-7}[\text{He}/\text{H}]_p $ : $ 0.1[\text{He}/\text{H}]_p $ : $ 10^{-7}[\text{He}/\text{H}]_p \}$.
    This assumes a mostly neutral atmosphere at the base of the hydrodynamic escape.
    \item For the stellar wind: $\{ 10^{-10} $ : $ \sim (1.0-[\text{He}/\text{H}]_\star) $ : $ 10^{-13}[\text{He}/\text{H}]_\star $ : $ 10^{-13}[\text{He}/\text{H}]_\star $ : $ 10^{-5}[\text{He}/\text{H}]_\star $ : $ [\text{He}/\text{H}]_\star \}$.
    This implies a nearly fully ionized stellar wind.
\end{itemize}
%The validity of such an HD model relies on the fact that the fluid remains within the collisional regime, i.e., the Knudsen number is smaller than unit.
%The latter compares the mean-free path of the species present in the fluid to the atmospheric scale height.

%Although only under approximations, we were able to confirm that this regime is obeyed.

%Note on non-collisional regimes (see Andrew's paragraph before 2.2)

We initialize our computational domain with a 1D velocity $\beta$ profile, taking the form $u_r = u_\infty(1-r^{-1})^\beta$, where $u_\infty$ is the terminal velocity of the outflow.
From mass conservation, we then initialize the density in the grid at the planetary boundary as $n(r)=n_0 r^{-2} u_0/u_r$.
%
%We begin our simulations with solemnly the launch of planetary material - the boundary condition at $-x$ is still inflow limiting.
%We allow the code to evolve the planetary wind for $N\approx8000$ steps, at which point the outflow is already quite close to a steady-state.
%We then turn on the tidal and Coriollis forces and let the simulation evolve for some more time (minimum of $N = 4000$ steps).
%
%Only after these steps, do we insert stellar winds by the outer boundary at $-x$.
%The boundary assumes a stellar wind with specific temperature, velocity and population densities, derived by an external 1D polytropic model with a nearly isothermal stellar wind - polytropic index $\alpha=1.05$.
%The stellar wind has the temperature and mass-loss rate fixed, and is not updated during runtime.
%We ultimately let this part evolve - depending on the simulation, the evolution takes more than $N=20000$ steps at all times, and $N\sim10^5$ steps for the most complex.
%Over the course of the evolution of the code, we have instructed it to print out the variables of interest, from hydrodynamic variables to reaction rates, heating and cooling values and $\tau_{\lambda}[\text{sp}]$.
%
%
We follow the same approach as \citet{21_Carolan}, in which we first allow the planetary outflow to relax in the grid, before turning on tidal and Coriolis forces.
Once this has relaxed, we inject the stellar wind through the $-x$ boundary until the system has reached a (quasi-)steady state.

To explore the effects of the stellar wind (SW) on the helium transit, we adopt several different stellar wind conditions for Star 1 and one wind condition for Star 2, as shown in Table \ref{tab:models_props}.
\af{
The stellar wind mass-loss rates vary between $5\times 10^{-16}$ and $1\times10^{-12} M_\odot \text{yr}^{-1}$, and we assume the wind is injected with the same velocity profile in all models.
At the boundary point of $\left( x=-48 R_p,\,y=z=0 \right)$, located at a distance of $\sim 3.3R_{\star,1}$ to the star, the wind, in the reference frame of the star, has a radial velocity of $29\, \text{km}\,\text{s}^{-1}$, which is obtained from assuming an isothermal stellar wind with temperature of $10^6$~K. Our grid is therefore in the acceleration zone of the stellar wind.
%From the planet's perspective, the stellar wind is injected in the exact same point with a radial velocity of $-24.7\, \text{km}\,\text{s}^{-1}$, and an absolute velocity of $62.7\, \text{km}\,\text{s}^{-1}$.
}
%
%
%The explored parameters have a main goal of being a proxy for stellar age.
%It is known that stars have a mass loss rate that decreases with age, though stellar wind evolution is understood by relating rotation with X-ray observations of stellar winds \citep[][]{03_Ivanova}, and rotation is related to age with the Skumanich law \citep[][]{72_Skumanich}.
%Several works have incorporated stellar winds in their simulations under this framework:
%\citet{15_Johnstone, 18_OFionnagain, 19_Carolan}.
%
%As a result, the models described in this work are not intended to track evolutionary progress, although they can help interpret possible helium line observations for distinct stellar systems, ranging from a possible young stellar host (cases \swthree\ and \swfour) to a weak companion (case \swone).
%
%
%
%This choice of parameters allows us to investigate the effects of distinct EUV fluxes and stellar wind strengths.
For reference, the current solar mass-loss rate is $\sim 2\times 10^{-14}M_\odot \text{yr}^{-1}$, thus our simulations have  $1/40\times$ (\swone), $10 \times$ (\swtwo\ and \swfour) or $50 \times$ (\swthree) the solar wind mass-loss rate, where the symbols inside the parenthesis are the IDs of our simulation runs (note that these numbers refer to mass-loss rates in units of $5\times 10^{-15} M_\odot ~ \textrm{yr}^{-1}$).

%Since the stellar wind temperature remains the same with different mass-loss rates, we are studying models where the injected velocity is slowly enhanced ($\sim 200 \text{km}\text{ s}^{-1}$) but have increasingly denser stellar winds.
%We assume a base planetary helium abundance of $10\%$ ($[\text{He}/\text{H}]_p = 0.1$) and a stellar wind helium abundance of $\sim 8\%$ ($[\text{He}/\text{H}]_\star = 0.078$), as values resembling the solar system.

We use the exoplanetary mass-loss rate $\dot{\mathcal{{M}}}_p$ as one of the evaluators to consider the simulations have reached a desired final steady-state.
It is calculated as:
\begin{equation}
    \dot{\mathcal{{M}}}_p = \oint_\Sigma \rho \vec{v} \cdot d\vec{\Sigma} \, ,
\end{equation}
where $\Sigma$ is a chosen closed \REV{spherical shell of radius $\mathbf{\mathrm{r}}$ }, and $\rho \vec{v}$ is the mass flux.
In Table \ref{tab:models_props} it is possible to see the planetary mass-loss rate for each different model, taken as an average between $\mathbf{\mathrm{r}}=2 \, R_p$ and $\mathbf{\mathrm{r}}=20\, R_p$.

\REV{To evaluate the impact of different resolution levels in the grid, we computed the mass-loss rate over closed shells of different radii.
While the mass-loss rate should remain constant with distance, we observe small discontinuous drops ($<0.5\%$) in $\dot{\mathcal{{M}}}_p$ slightly after $\mathbf{\mathrm{r}}=3R_p$, $\mathbf{\mathrm{r}}=5R_p$ and $\mathbf{\mathrm{r}}=7.5R_p$, which are corresponding positions of change in cell resolution. %%Testing a grid that is one level of refinement less resolved in all the regions described in Section 2.4 (except for the inner highly resolved shell), we find a difference of $1.6\%$ in the averaged $\dot{\mathcal{{M}}}_p$ value.
%Thus, we are confident that the employed grid is not impacting converged solutions with propagated resolution artifacts.
}

\begin{table*}
\centering
\caption{Summary of the adopted parameters of our models, as well as global output values. The columns related to the input parameters are: the model ID, the host star ID, the stellar host radiation fluxes in four different energy bands (see text), the stellar wind mass-loss rate ($\dot{\mathcal{M}}_\star$) and the size of the Hill sphere ($r_\text{H}$). The columns related to the output of our simulations are: the planetary mass-loss rate ($\dot{\mathcal{M}}_p$) and the distance to the flow-flow interaction region ($\mathrm{d}_\text{shock}$). }
\label{tab:models_props}
\begin{tabular}{c || cccccccccc}
\hline
Model & Star & $\mathcal{F}_{\mathrm{X-ray}}$ & $\mathcal{F}_{\text{hEUV}}$ & $\mathcal{F}_{\text{sEUV}}$ & $\mathcal{F}_{\text{mUV}}$ & $\dot{\mathcal{M}}_\star$ & $r_\text{H}$ & & $\dot{\mathcal{M}}_p$ & $\mathrm{d}_\text{shock}$
\\
  &  & \multicolumn{4}{c}{($10^{3}\, \text{erg} \, \text{cm}^{-2} \, \text{s}^{-1}$)} & ($5\times10^{-15} M_\odot\, \text{yr}^{-1}$) & ($R_p$) & & ($10^{10}\text{g}\text{s}^{-1}$) & ($R_p$)
  \\
\hline
\hline
\swone\ & \multirow{5}{0.65cm}{Star 1} & \multirow{5}{0.5cm}{$0.15$} & \multirow{5}{0.5cm}{$1.0$} & \multirow{5}{0.5cm}{$0.8$} & \multirow{5}{0.5cm}{$50$} & $0.1$ & \multirow{5}{0.4cm}{6.2} & & $1.1$ & $12.0$ 
\\
\texttt{S1.6}\ &   &   &   &   &   & $1.6$ & & & $1.1 $ & $10.0$ 
\\
\texttt{S8.0}\ &   &   &   &   &   & $8.0$ & & & $0.9 $ & $2.9$
\\
\swtwo\ &   &   &   &   &   & $40$ & & & $ 1.0 $ & $2.0$ 
\\
\swthree\ &   &   &   &   &   & $200$ & & & $ 1.0 $ & $1.5$ 
\\
\hline
\swfour\ & Star 2 & $18$ & $15$ & $22$ & $4300$ & $40$ & $5.9$ & & $24.6 $ & $33.0$ \\
\hline
\hline
\end{tabular}
\end{table*}

\begin{figure*}
    \centering
        \includegraphics[width=\linewidth]{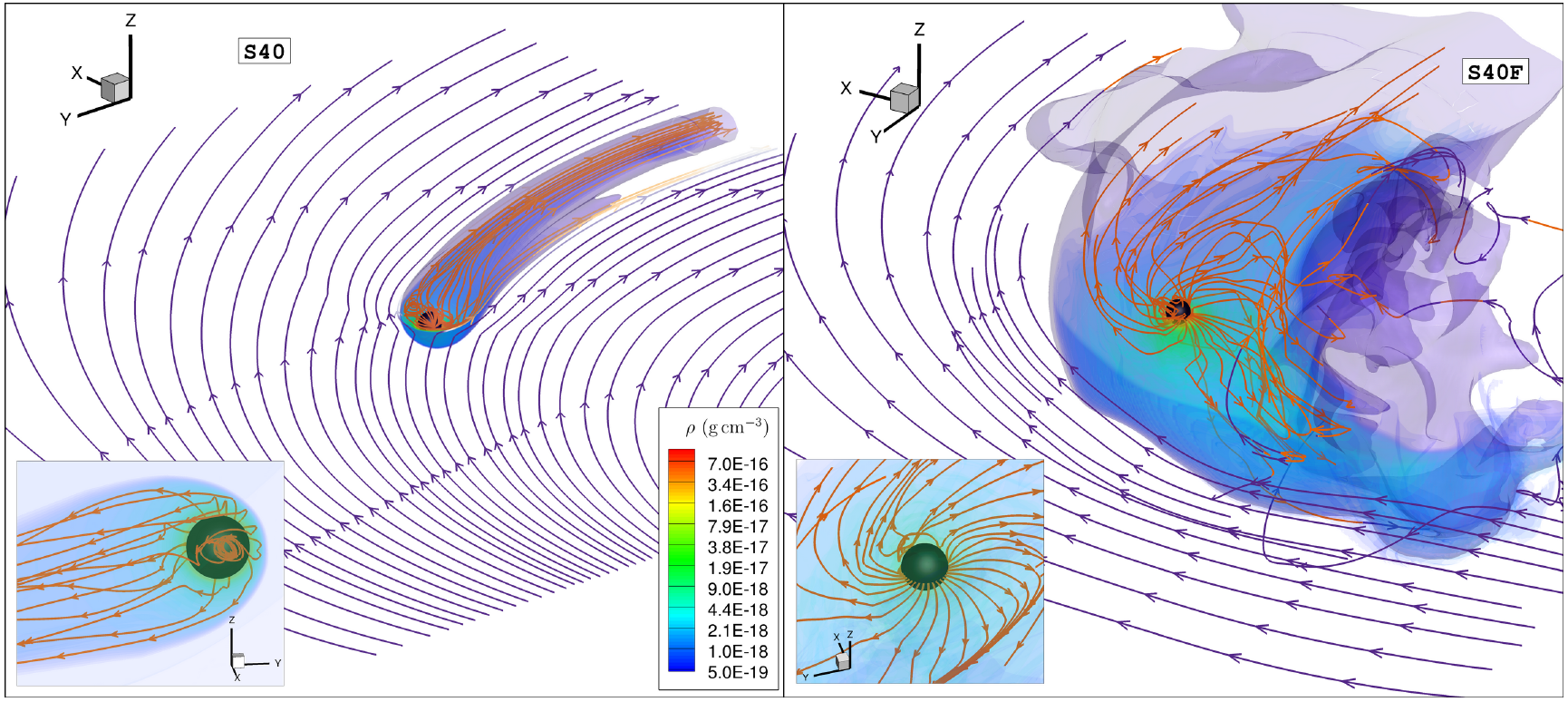}
    \REVcaption{3D view of the modeled evaporating planetary atmosphere \swtwo\ (left) and \swfour\ (right), which have the same stellar wind mass-loss rates, but different stellar XUV fluxes.
    %The colors show the ion number density in the orbital plane $z=0$ and the velocity streamlines are colored according to the absolute velocity.
    %Within (beyond) the isosurface, the fluid is mainly composed of planetary (stellar wind) material.
    %\REV{NEED to COREECT}
    \REV{We show the exoplanet as the black sphere, with total density ($\rho$) displayed in several isosurfaces - these have been cut for $z>0$ in the full view panels - and velocity streamlines are colored by association to an initial planetary outflow (orange) or stellar (blue) wind. 
    The dark blue outer closed isosurface delimits where the gas is made up of mostly planetary material. 
    The insets show a closer view of the planet.}
    }
    \label{fig:3D}
\end{figure*}

%\subsection{HD Outputs}

\section{Results}
\label{section4}

%In this subsection, we present some of the outputs of the different simulated cases .
%We typically run our models for at least $60$ thousand iterations until the simulations reach steady state.
The typical morphology that we find in such cases (\texttt{S8.0}, \swtwo, \swthree) has a comet-tail like structure, where most of the up-orbit outflow is redirected towards the orbital path behind the planet.
However, we also have cases where the outflow and SW interaction behaves cyclically - in a quasi-steady state. These are \swone, \texttt{S1.6} and \swfour.
This is because the extension of photoevaporated material into a frontward arm is subject to turbulence in the flow-flow interaction (e.g., see right panel of Figure \ref{fig:3D}; movies showing the progress of simulations can be found \href{https://doi.org/10.5281/zenodo.21455405}{here}.).

%\REV{NEED TO CORRECT}
\af{
Figure \ref{fig:3D} shows 3D views of \swtwo\ and \swfour, where it is possible to distinguish the different outflow geometries that arise from the interaction between a planet's outflow with two different stellar hosts, but sharing the same stellar wind conditions.
}
\REV{The planet is shown as the central black sphere, with velocity streamlines colored by their object of origin: orange lines for planetary material and blue lines for the stellar wind.
Density is displayed by several isosurfaces that have been cut for $z>0$ in the full view panel, while the insets show a close-in look to the launch of the outflow without this blanking condition.
}
The \REV{dark blue} closed isosurface is related to the passive scalar variable.
This delimits the region where most of the material is originally from the escaping planetary atmosphere (within the isosurface), and beyond which, the material is predominantly stellar wind.

%The planet is shown as the central black sphere, with streamlines colored by the absolute velocity of the flow, and the $xy$ plane is color coded by the number density of ions $n_\text{ion}=n_{\text{H}^+}+n_{\text{He}^+}+n_{\text{He}^{++}}$.
%The isosurface is related to a passive scalar variable, which delimits the region where most of the material is originally from the escaping planetary atmosphere (within the isosurface), and beyond which, the material is predominantly stellar wind.
%The color of streamlines indicates that planetary gas is an order of magnitude slower than the velocity of the stellar wind.

Note that the extension of the planetary outflow arises naturally in our 3D models.
Such a feature is not possible to obtain through 1D models, where most of the times the outflow extension is defined \textit{a priori} (for example, stopping at the Hill radius, as in \citealt{Caldiroli--2021}, or at a fixed distance, as in \citealt{23_Allan}).
\af{
It is possible to associate different morphologies of hydrodynamic escape with typical distances embedded in each star-planet system \citep[][]{15_Matsakos}.
Although this is beyond the scope of this work, we see that these characteristic lengths - like the Hill radius ($r_\text{H}$) and the distance to flow-flow interaction ($d_\text{shock}$) - can entail quite distinct synthetic spectra, highlighting the need of 3D models in computing the extension of escaping atmospheres.
%Since the Hill distance is independent of the stellar wind property, it will be such characteristic lengths that will distinguish outflow geometries \citep[][]{15_Matsakos}, highlighting the need of 3D models in computing the extension of escaping atmospheres.
}
\REV{Given that some simulations show planetary outflows extending to the boundaries before reaching steady-state, we have checked for boundary effects that could artificially impact the system's dynamics. 
Our analysis confirms that no erratic behavior is taking place, both in total energy or in the velocity profiles.
%Despite this, all boundaries have the lowest level of cell resolution and our 'inflow limited' boundary condition ensures a typical outflow - similar to \citet{19_McCann}.
%We present most outputs in the following sections in the $z=0$ plane with shorter domains than the real grid, simply for better visualization purposes.
}

%\subsection{First set of models: K-dwarf star}
\subsection{Effects of stellar wind on the Helium population}

\begin{figure*}
    %\captionsetup{labelformat = newfig}
    \centering
        \includegraphics[width=\linewidth]{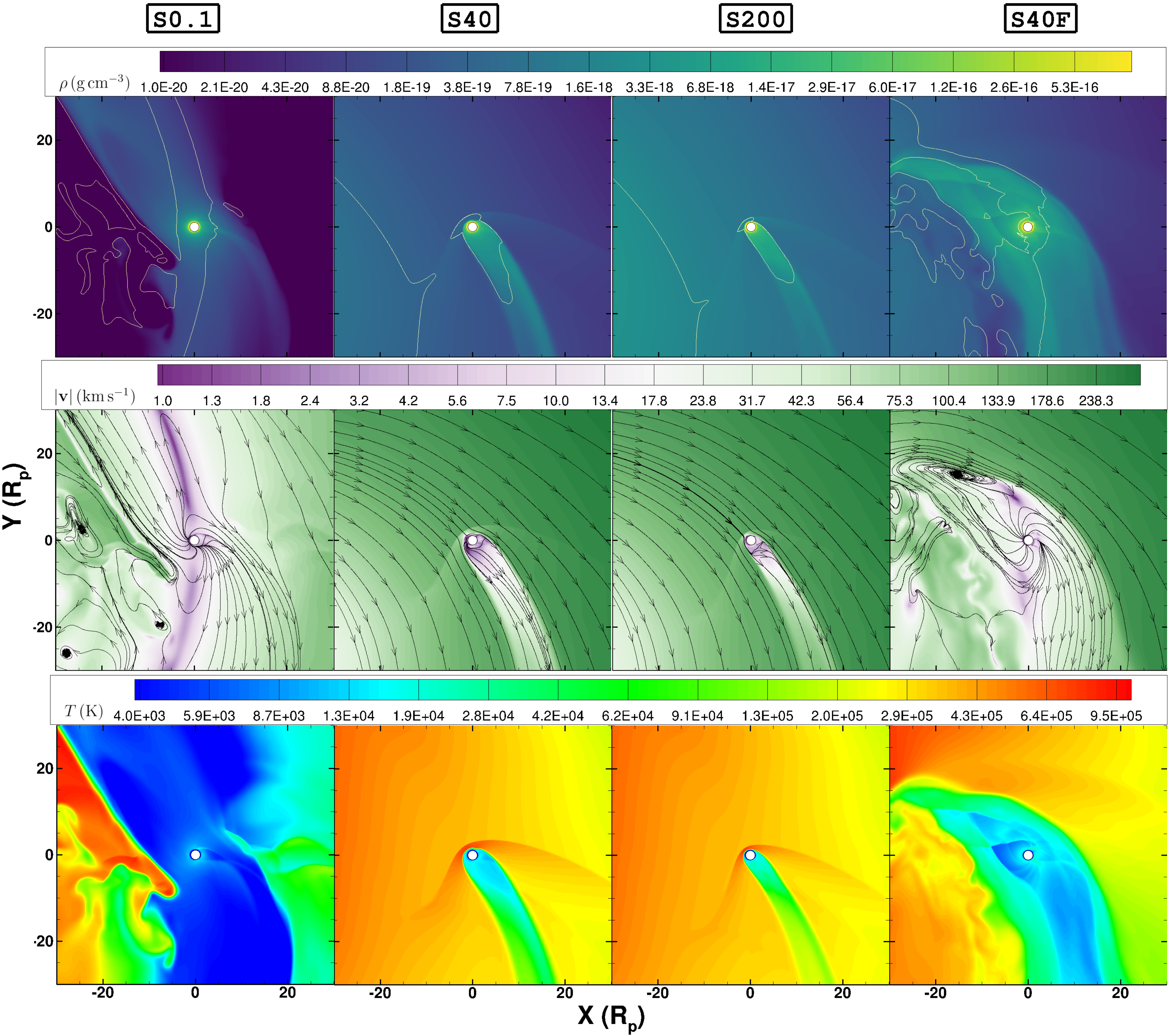}
    \REVcaption{
    2D view of the density (top), total velocity (middle) and temperature (bottom) of models \swone, \swtwo, \swthree\ and \swfour. 
    %The pink line shows the sonic surface and the streamlines represent the velocity.
    The \REV{pale yellow} line shows the sonic surface and the streamlines represent the velocity.
    }
    \label{fig:2D-HDvars}
\end{figure*}

For the purpose of our discussion, we will focus on the cases \swone, \swtwo, \swthree, which assume the same stellar XUV flux, but varying stellar wind mass-loss rates ($1/40$, $10$ and $50$ times the solar values).
Figure \ref{fig:2D-HDvars} shows contours of fluid density (top), absolute velocity (middle) and temperature (bottom panels) in the orbital ($z=0$) plane.
There are also other informative lines, such as the \REV{pale yellow} line in the upper plots, which traces the sonic surface at Mach number $\mathrm{Ma}=1$, and velocity streamlines in the middle panels.
We see that the strengthening of SW causes the volume occupied by the planetary material to decrease, as expected \citep[e.g.,][]{21_a_Carolan}.
%From the top plots, we see the inserted SW growing in density according to the model, while the planetary material evolves inside the shocked zone.
Model \swone\ shows the largest volume occupied by the planet's escaping material, with a clear presence ahead of orbit, while \swtwo\ and \swthree\ have the outflowing gas funneled into a comet-like tail structure.
In the middle panels, the absolute velocity of the planetary gas indicates a turbulent structure when interacting with a low SW strength (\swone) similarly to the case \swfour\ discussed before.
As for models \swtwo\ and \swthree, the escaping material is strongly advected to the nightside and behind orbit.
The planetary material reaches typical velocities of $\sim 20\ \text{km}\,\text{s}^{-1}$ at the tail of \swone\ and up to $\sim 90\ \text{km}\,\text{s}^{-1}$ for \swthree.
Finally, the bottom panels show the temperature structures, where we can distinguish shocks by jumps in temperature. 
We emphasize the different temperatures of the outflowing gas (shades of blue and cyan) in the different tails behind orbit: \swone\ outflow is mostly cold ($T\sim 5\times10^3\ \text{K}$) and does not go over $1.2 \times10^4 \text{K}$, while \swthree\ is above $10^4\ \text{K}$ nearly over the entire volume.

The shock is formed at the point where there is pressure balance between the stellar wind and the planetary outflow. 
We take the distance of the bow shock, $\mathrm{d}_\text{shock}$, to be where this condition takes place the furthest in the planet's dayside (stated in Table \ref{tab:models_props}).
We can find $\mathrm{d}_\text{shock}$ bigger than the sonic point ($|\mathbf{v}|= c_s$ at $r_\text{sonic}$) in case \swone\ (and \swfour), where the flow is transonic; and when this condition fails, the outflow is considered a breeze.
Case \swone, for example, develops an extended front-arm of cold gas that is highly influenced by orbital effects, since the stellar wind wraps around this stream-like structure.
Thus, material starting its launch on the dayside of the exoplanet, crosses the sonic surface ($r_\text{sonic} = 3.9 \, R_p$) and is pushed ahead of orbit, becoming a transonic flow.
Models \swtwo\ and \swthree\ have the dayside-launched gas quickly advected to the nightside.
Thus, this material starts its escape in a breeze condition and finds a transonic solution on the rather stable comet-like tail structure.

%In summary, we have a transonic atmospheric escape for model \swone\ and dayside breezes for \swtwo\ and \swthree\ that become transonic in the nightside.

Figure \ref{fig:2D-energy} introduces additional information such as total volumetric heating rate, total volumetric cooling rate and PdV work.
The volumetric heating rate has contributions from the flux heating, which is the release of excess energy from stellar flux photoionization, and recombination heating, from photons ($\lambda_\alpha$) originated in the recombination process $\alpha_\text{A-B}$ that add to the existing photoionization.
When integrating heating over volume, $\mathcal{P}_\mathcal{H}$, we find recombination to be the leading heat source for all models of Star 1.
The nightside close to the planet only has heating contributions from the recombination photons.
On the other hand, we have 12 cooling sources - Table \ref{tab:cooling_processes} - with hydrogen recombination cooling as the leading coolant inside the planetary outflow.
%Notice how the total cooling increases with increasing stellar wind ram pressures inside the volume of planetary material.
In the middle row, we observe that the total cooling increases in the come-like tail region as the stellar wind strength rises from left to right.
%By comparison, although both \swtwo\ and \swthree\ have similar comet-like tails of outflowing material, the hydrodynamics within these are quite distinct as the volume of material decreases with an increase in $|\mathcal{H}-\mathcal{C}|$.
%The prevailing cooling terms are from recombination and collisional excitation, mostly by means of hydrogen.
From an increase in the stellar mass-loss rate, we see that pressure balance takes place closer to the planet, with steeper gradients of temperature.
This impacts the underlying properties of the outflow, originating contrasting differences in the population results.
For example, both \swtwo\ and \swthree\ show similar comet-like tails of outflowing material.
But as a consequence of decreasing the available volume for heating, increasing cooling and PdV work (Equation (\ref{EQ--energy})), it originates a tail with enhanced gas temperatures.

%An additional note can be included: after evolving from the initial shock between fluids, as some material was advected to the southwestern corner ($-x$ and $-y$) and is not blown away to the boundaries.

%\aav{in the top panels, we see that the volume occupied by the atmosphere reduces as we increase the sw mass-loss rate. We also see that the planetary material remains subsonic in the dayside of SW2 and SW3, [talk about breeze if you want], and instead is transonic in the nightside, where we see the formation of a comet-like tail. In the middle panels, we see that planetary outflow reaches velocities of XXX km/s in all cases. We also see a more turbulent structure in SW1, which has the extended frontarm. Finally, the bottom panels show the temperature structure, where we see a jump in temperature contrast at the position of the shock. }

\begin{figure*}
    \centering
        \includegraphics[width=\linewidth]{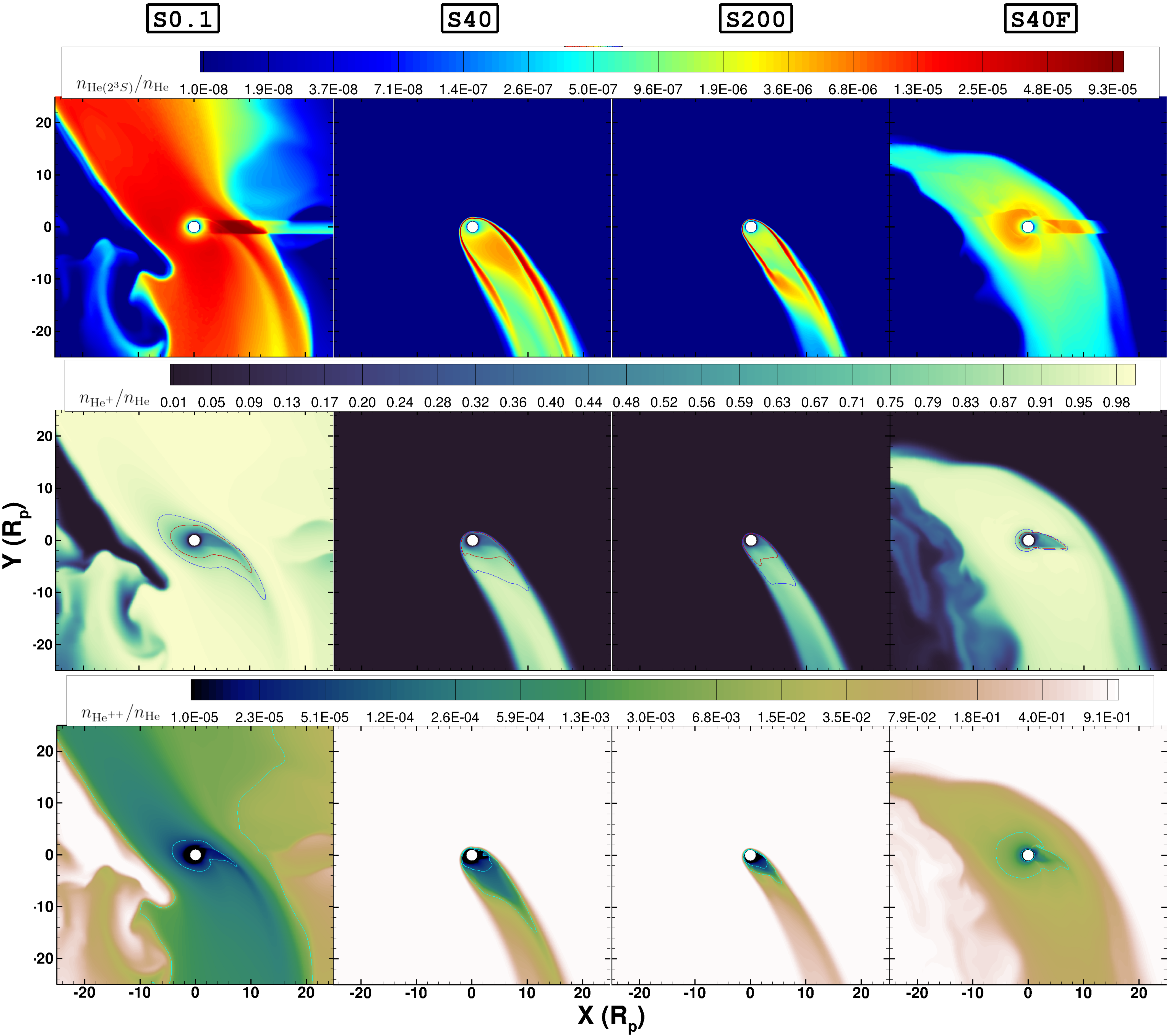}
    \REVcaption{
    2D view in the orbital plane $z=0$ of the fraction of the He species: triplet ($\text{He}(2^3S)$), ionized helium ($\text{He}^+$) and doubly ionized helium ($\text{He}^{++}$) for the top, middle and bottom rows, respectively.
    Each column corresponds to a distinct model.
    The red line marks $n_\text{ion}=0.9$, the blue line marks $n_\text{He-ion}=0.9$ and the cyan lines mark where $n_{\text{He}^{++}}/n_{\text{He}^+}$ is $10^{-4}$ and $0.01$ for the inner and outer lines, respectively.
    }
    \label{fig:2D-He-species}
\end{figure*}

Figure \ref{fig:2D-He-species} shows the ratios of He species by the following order: helium triplet (top panels), ionized helium (middle) and doubly ionized helium (bottom).
These were calculated as $n_\text{sp}/n_\text{He}$, where $n_\text{He} = n_{\text{He} (1^1S)} + n_{\text{He} (2^3S)} + n_{\text{He}^+} + n_{\text{He}^{++}} $.
The first row displays different fraction values and structures, indicating that the species $\text{He}(2^3S)$ is influenced by stellar winds. 
In the case of \swone, we see that the species $\text{He}(2^3S)$ in the planetary outflow is enhanced and extends over large distances, both in front and trailing the planet.
Focusing on the central region of the outflows ($r \lesssim 7.5 \, R_p$) and tail, we find relative triplet fractions of the order $\sim 10^{-5}$ in \swone, $\sim 5 \times 10^{-6}$ for \swtwo\ and \REV{$\lesssim 10^{-6}$} for \swthree.
Models \swtwo\ and \swthree\ do not reach such high fractions of triplet state in their comet-like tail, except for the flank region of the shock. As we will discuss below (Section \ref{section6}), this has consequences for helium transits.
In these specific flank regions at the shock, He triplet fractions can reach values of order $\gtrsim 10^{-5}$ for \swtwo\ and \swthree.
\af{
The middle plots of Figure \ref{fig:2D-He-species} show the first ionized state of He.
The red line marks where the total gas ionization level is  $90\%$, while the blue line  marks where the helium gas is $90\%$ ionized.
}
The state $\text{He}^+$ is the dominant state of He in the evaporating atmosphere, except by state $\text{He}(1^1S)$ close to the base of the launch ($\lesssim 2\, R_p$) and in the nightside.
The lower row displays the second ionized state $\text{He}^{++}$, the state that makes up most of the He in the hot stellar wind that shocks with a colder, denser and less ionized gas.
\af{
There, two cyan lines mark where the ratio of doubly ionized to ionized He $n_{\text{He}^{++}}/n_{\text{He}^+}$ is $0.01\%$ (inner closed line) and $1.0\%$ (outer closed line).
The planetary species $\text{He}^{++}$ is mostly absent in the cold planetary outflow, only becoming enhanced in highly turbulent regions that mix with the stellar $\text{He}^{++}$.
}

%For case \swone, the double ionized state is mostly present in the turbulent regions, as the optical depth is low, in relative fractions of $\sim 0.1$.
%In models \swtwo\ and \swthree\ we find that the tail reaches values of $\gtrsim 0.1$, compensated by $\text{He}^+$ that becomes ionized by X-rays.

%The higher the mass-loss rate of the stellar wind, the quicker we see planetary gas reaching the doubly-ionized state of He.
\begin{figure*}
    \centering
        \includegraphics[width=0.75\linewidth]{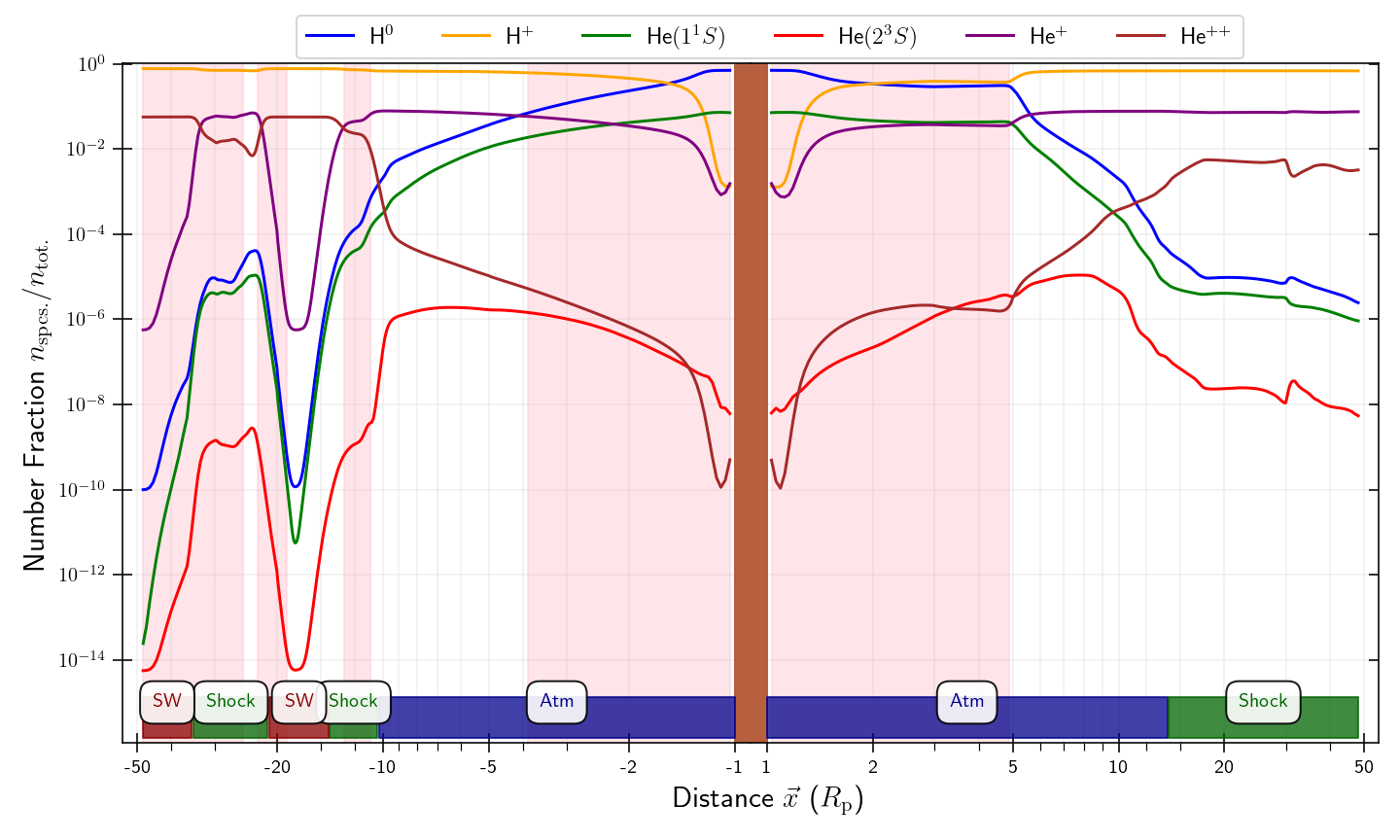}
    \REVcaption{Comparison between ratio of individual species to total number of species, following the line connecting the star to the planet ($z=0$, $y=0$), for the case \swone.
    The central brownish region represents the planetary surface.
    The filled pink regions target the space where the flow is subsonic.
    Above the x-axis are displayed the 3 distinct regions of the model: stellar wind (`SW'), the shock and the planetary atmosphere that is outflowing (`Atm').}
    \label{fig:1D-fraction-species}
\end{figure*}

Our findings reveal, comparatively, that for stronger stellar winds, the tail-like outflow shows decreasing helium triplet fractions, only being specifically enhanced in the flank zone of the shock.
Close to the exoplanet ($r < 2.5R_p $), the fractions of $\text{He}^+$ and $\text{He}^{++}$ are low, since the optical depth for photoionization is high and temperature is low.
In a cold planetary outflow ($T\lesssim 1.5\times10^{4}\, \text{K} $) at larger distances, the $\text{He}^+$ species dominate, quickly reaching $n_{\text{He}^+}/n_{\text{He}}=0.9$, mostly by photoionization of the singlet state.
Moreover, as seen in the case of \swthree, the highly compressed atmosphere in the tail promotes the $\text{He}^{++}$ state to the second most abundant state with $n_{\text{He}^{++}}/n_{\text{He}} \sim 0.1$.
We remind the reader that this is only relative to He species, as the ionized H is still the overall dominant gas particle.
%\aav{, only being surpassed by the XXXX. [PS make sure that you tell the reader that you are only focusing on helium, as I would expect that protons are still the dominant state, right?]}.

To deepen the analysis on the population of all species, Figure \ref{fig:1D-fraction-species} shows the fraction of species compared to the total number of species along the $x$ axis (i.e., the star-planet line) for \swone.
We specify the extent of the stellar wind (SW), the shock and the planetary atmosphere (Atm).
We see that, far from the planet, the population of species varies sharply due to the stellar wind. For close-in regions, there is an asymmetry of populations for dayside ($-x$) and nightside ($+x$).
%Our density ratios on the dayside are similar to the 1D models of \citet{23_Allan} in the region with $r \lesssim r_\text{H} \sim 6R_p$.
% We also see an enhancement of the triplet species from $r=4R_p$ to $7R_p$ in the nightside of the planet. This is the region where photoionization does not occur, due to the planet obscuring the stellar flux.
% This increase comes from the increase of $\text{He}^+$, the primary reservoir for $\text{He}(2^3S)$ by recombination.
Given that in the nightside of the planet stellar photons cannot ionize helium, intuitively, we would have expected a low fraction of $\text{He}^+$ and $\text{He}^{++}$ in the nightside, but this is not what we found.
These ions are created in atmospheric regions that are unshielded to stellar photonionisation (e.g., from the $+y$ direction) and subsequently advected to the nightside.
An interesting consequence of this is that, because $\text{He}^+$ is the primary reservoir of $\text{He}(2^3S)$ via recombination ($\alpha_{\text{B}}[\text{He}(2^3S)]$), we also see an enhancement of the triplet species in the nightside of the planet. 
We discuss the importance of considering the advection terms in 3D models in Appendix \ref{Section_Advectionless}.

\begin{figure*}
    \centering
        \includegraphics[width=\linewidth]{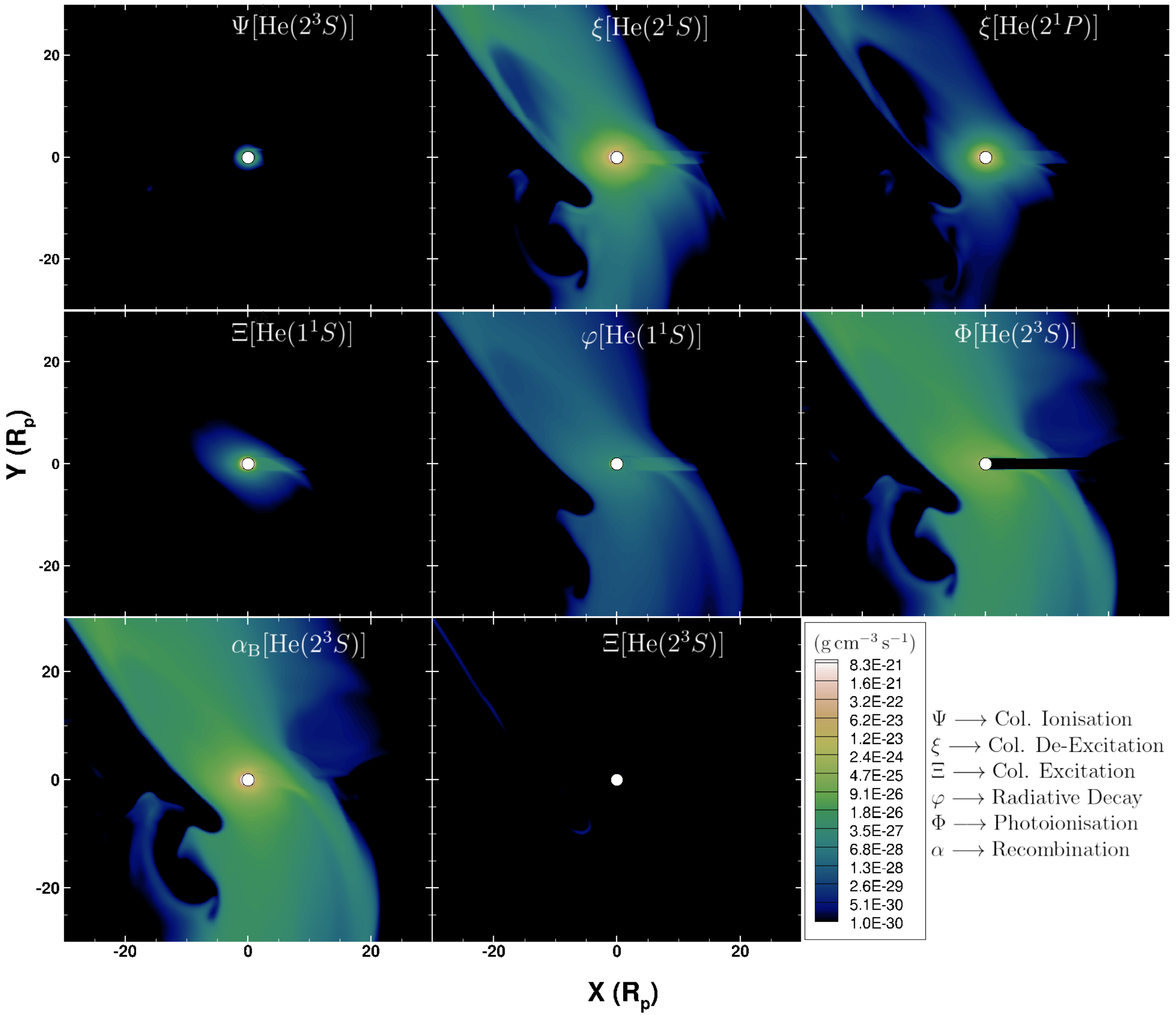}
    \REVcaption{2D view in the orbital plane $z=0$ of all contributing reactions rates to the population of He triplet for the model \swone. 
    Top and middle rows: annihilating rates.
    Bottom row: populating rates.
    Note that the contribution of photoionization ($\Phi$) is incomplete, as photons that originate from the recombination reaction will also contribute.
    Figures \ref{fig:2D-SW2-All-triplet-rates}, \ref{fig:2D-SW3-All-triplet-rates} and \ref{fig:2D-SW4-All-triplet-rates} show the equivalent plots for models \swtwo, \swthree\ and \swfour.
    Figure \ref{fig:2D-SWs-Net-triplet-rates} shows the net rate.
    }
    \label{fig:2D-SW1-All-triplet-rates}
\end{figure*}

The extensive reaction chain underlying this model makes the analysis of each species somewhat lengthy, so we focus mainly on the helium triplet here.
Figure \ref{fig:2D-SW1-All-triplet-rates} depicts the contributing rates for the triplet, in the $xy$ plane, for the same model as before, \swone. For \swtwo\ and \swthree, see Figures \ref{fig:2D-SW2-All-triplet-rates} and \ref{fig:2D-SW3-All-triplet-rates}, respectively.
These plots aid our understanding on large scale spatial contributions of each reaction (de)populating the triplet.
\af{
We see that $\text{He}(2^3S)$ balances itself mainly between recombination $\alpha_{\text{B}}[\text{He}(2^3S)]$ and collisional de-excitation $\xi[\text{He}(2^1S)]$.
Moreover, close to the planet, some reaction rates barely contribute to the net rate of the species, as seen in several 1D works \citep[][]{23_Allan, 25_Taylor, 25_Munoz}.
However, our 3D models show that they become important in the flow-flow interaction zone.
}
%There are reaction rates that become enhanced in specific structures: for example, the collisional de-excitation of He singlet into He triplet by means of a free electron ($\Xi[\text{He}(2^3S)]$, last panel) has a negligible contribution throughout most planetary outflow.
%\af{
%However, in regions of flow-flow interaction and at the flank of the shock, the latter collisional rate ($\Xi[\text{He}(2^3S)]$) reaches values that overcome recombination rates, enhanced by stronger stellar winds.
%It is possible to see this when comparing the last plot of Figure \ref{fig:2D-SW1-All-triplet-rates} with the same one in Figures \ref{fig:2D-SW2-All-triplet-rates} and \ref{fig:2D-SW3-All-triplet-rates}
%}
%Contrastingly, the rate of collisional de-excitation of He triplet into He singlet by means of a free $\text{H}_0$ ($\Xi[\text{He}(1^1S)]$, first panel in the middle row) is only relevant for distances close to the planet, as this region contains most neutral H species.

%There are reaction rates that become enhanced in specific structures: for example, the collisional de-excitation of He singlet into He triplet by means of a free electron ($\Xi[\text{He}(2^3S)]$, last panel) has a negligible contribution throughout most planetary outflow.
%\af{
%However, in regions of flow-flow interaction and at the flank of the shock, the latter collisional rate ($\Xi[\text{He}(2^3S)]$) reaches values that overcome recombination rates, enhanced by stronger stellar winds.
%It is possible to see this when comparing the last plot of Figure \ref{fig:2D-SW1-All-triplet-rates} with the same one in Figures \ref{fig:2D-SW2-All-triplet-rates} and \ref{fig:2D-SW3-All-triplet-rates}
%}
%Contrastingly, the rate of collisional de-excitation of He triplet into He singlet by means of a free $\text{H}_0$ ($\Xi[\text{He}(1^1S)]$, first panel in the middle row) is only relevant for distances close to the planet, as this region contains most neutral H species.

%Moreover, close to the planet, some reaction rates barely contribute to the net rate of the species, as seen in several 1D works \citep[][]{23_Allan, 25_Taylor, 25_Munoz}.
%However, our 3D models show that they become important in the flow-flow interaction zone.
\af{
In a case of a very weak stellar wind (Figure \ref{fig:2D-SW1-All-triplet-rates}), triplet-destructive rates like collisional ionization $\Psi[\text{He}(2^3S)]$ or the populating rate, collisional excitation $\Xi[\text{He}(2^3S)]$, add insignificantly to the triplet's net rate.
Whereas for a mild to strong stellar wind interacting with the photoevaporated gas - as we can see from Figures \ref{fig:2D-SW2-All-triplet-rates} and \ref{fig:2D-SW3-All-triplet-rates} - these rates can become extremely relevant at the shock zone.
This is explained by an increase in electron number density being mixed with the gas and local temperature increase.
Contrastingly, the rate of collisional de-excitation of He triplet into He singlet by means of a free $\text{H}_0$ ($\Xi[\text{He}(1^1S)]$, first panel in the middle row) is only relevant for distances close to the planet, as this region contains most neutral H species.
}
Thus, we have shown that the seemingly non-contributing rates for defining the population along the cold planetary outflow turn out to be relevant when stellar wind is present, as the hot mixed region enhances these rates.
This region becomes a rich zone of population-chemistry, as is also depicted in Figure \ref{fig:2D-He-species}, where the fractions of He species are prone to change strongly.
\af{
In Figures  \ref{fig:2D-SWs-Net-triplet-rates} and \ref{fig:1D-net-rates} we plot the net rate of species, which is defined as the summed contribution of the sources minus the sinks for each species $\mathcal{R}[\text{sp}] = \left[ \mathcal{S} - \mathfrak{S} \right]_\text{sp}$.
}
\subsection{Effects of stellar flux on the Helium population}

We now move on to the atmospheric model of a more irradiated planet. 
In \swfour, we use a high energy spectrum and a stellar wind of 10 times the solar wind mass-loss rate, the same value as adopted in \swtwo. 
Therefore, by comparing \swtwo\ and \swfour, we gain insight in the helium population for the case where only the XUV flux varies.

The rightmost panels in Figure \ref{fig:2D-HDvars} show the density, absolute velocity and temperature of \swfour, where we see a dayside transonic escape with several shock structures inside the planetary outflow.
The immense photoionizing power derived from an extremely high XUV flux from the star results in a high mass-loss rate (Table \ref{tab:models_props}), and consequently, a strong ram pressure that leads to an extended and quasi-steady arm of material ahead of orbit.
By comparing with the model \swtwo, the planet's mass-loss rate $\dot{\mathcal{M}_p}$ increases $\sim 25$ times.
For this system, the sonic radius ($r_\text{sonic} \approx2.1 \, R_p$) is $\sim 1.8 \times$ closer to the planet, also smaller than the largely unaltered Hill radius.
Thus, the outflowing gas extends ahead of orbit through strong acceleration and heating.
In contrast with the turbulent \swone\ model, this frontward arm evolves somewhat a cyclically -- increasing and decreasing its extension towards the star.

The right panels in Figure \ref{fig:2D-energy} show that a large volume of material exhibits strong heating.
We compute the volume integral of the volumetric heating and cooling rates, starting at the planetary radius up to a distance $r_f$, which provides us the total heating and cooling powers ($\mathcal{P}_\mathcal{H} ,\, \mathcal{P}_\mathcal{C}$ respectively).
The results at $r_f=3\, R_p$ show powers of $\mathcal{P}_\mathcal{H} = 4.5\times10^{24}\, \text{erg} \, \text{s}^{-1}$ and $\mathcal{P}_\mathcal{C} = 7\times10^{23}\, \text{erg} \, \text{s}^{-1}$.
Comparing with the previous set of models, cooling showed similar values of total power, while heating showed $\mathcal{P}_\mathcal{H} = 1.8\times10^{24}\, \text{erg} \, \text{s}^{-1}$, mainly due to recombination.
The \swfour-dominant heating channels are the photoionization of $\text{H}_0$ by sEUV photons, or $\text{He}^+$ by X-rays close to the shock zones.
Thus, there is a clear stellar flux heating dominance in \swfour, in contrast with a shared heating between recombination (main contributor) and stellar flux in \swone\ to \swthree.
%Given that models \swtwo\ and \swfour\ have the same stellar wind and planetary properties, the reason for the difference in heating is solely due to an increase irradiation flux.

Figure \ref{fig:2D-SW4-All-triplet-rates} shows the triplet reaction rates and the bottom right panel in Figure \ref{fig:2D-SWs-Net-triplet-rates} shows the net rate of the same species. There is no significant region of triplet enhancement, even though all rates now display higher values.
Spatially close to the planet, we have recombination $\alpha_{\text{B}}[\text{He}(2^3S)]$ and collisional de-excitation $\xi[\text{He}(2^1S)]$ as the leading sources and sinks of triplet, respectively.
However, at $r\sim 3\, R_p$ in the dayside of the planet, the main sink becomes photoionization $\Phi [\text{He}(2^3S)]$, as mid-UV photons are $\sim 86 \times$ higher than the previous set of models.
%This contrasts with the previous set of models, where we only see in \swone\ this same replacement of rates at further distances ($r\sim5R_p$).
As we are now dealing with stronger fluxes, which facilitates the ionization of neutral $\text{He}$, the dayside point at which gas becomes mainly ionized is now at $r\sim 1.3\, R_p$.
For that reason, we can see in the right panels of Figure \ref{fig:2D-He-species} that both ionized He species make up most of the outflowing volume of material.
The fraction of triplet at close distances to the exoplanet is around the order of \REV{$4 \times 10^{-6}$}, but for $r>10\, R_p$ ahead of orbit the density of triplet drops to \REV{$\lesssim \times 10^{-6}$}.
\af{
The flux of X-ray photons is also $\sim 121$ times higher, which enhances photoionization in this channel, which strongly ionises $\text{He}^+$ to $\text{He}^{++}$.
%The enhancement of $\text{He}^+$ and $\text{He}^{++}$, as we see in Figure \ref{fig:2D-He-species} from the ratio of ionized to doubly ionized $n_{\text{He}^{+}}/n_{\text{He}^{++}} \sim 95/5$.
The cyan line countours on the Figure \ref{fig:2D-He-species} support this: the outer closed contour that marks $n_{\text{He}^{++}}/n_{\text{He}^+}=1\%$ is found at the shock of \swone, but well inside the planetary outflow for a stronger flux case.
When comparing the fraction of $\text{He}^{++}$ in the tail between \swtwo\ and \swfour, we find respectively $n_{\text{He}^{++}}/n_{\text{He}} \lesssim 10^{-3}$ compared with $n_{\text{He}^{++}}/n_{\text{He}} \lesssim 10^{-1}$.
}

\section{Synthetic Spectra}
\label{section5}

\REV{\subsection{Radiation Transfer Setup}
}

After reaching a final state, we run each simulation output through a ray-tracing model to retrieve synthetic transmission spectra. We use the ray-tracing model described in \citet{21_a_Carolan}, with the line parameters of the helium triplet line, which is composed of three individual lines, with line-center wavelengths of $\lambda_1 = 10829.09114$  \r{A}, $\lambda_2 = 10830.25010$ \r{A} and $\lambda_3 = 10830.33977$ \r{A} in air. 
The Einstein coefficient is $1.0216 \times 10^{7} \text{ s}^{-1}$ for all lines; their corresponding oscillator strengths are $0.059902$, $0.17974$, $0.29958$ \citep[extracted from the NIST database,][]{24_NIST}.
The He I line transit is the added contribution of each of these individual line transits.

We assume a uniform specific intensity over the stellar disc at a given frequency (i.e., no limb darkening) and the line profiles are calculated for a null impact parameter.
With the model parameters, we obtain a transit time of $t_\text{T} = 3.22 \text{h}$ for \swone, \swtwo, \swthree\ and $t_\text{T} = 3.44 \text{h}$ for \swfour.

\begin{figure*}
    \centering
        \includegraphics[width=\linewidth]{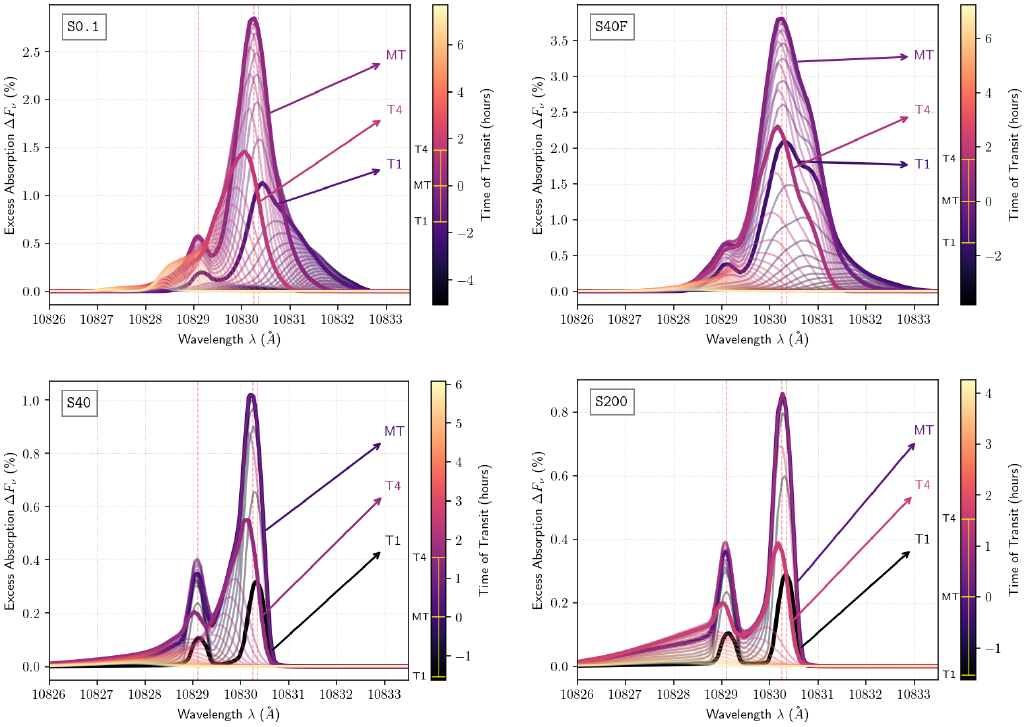}
    \REVcaption{Excess absorption of the He I line for all our simulations at different times.
    The solid lines corresponding to first contact (`T1'), mid-transit ($t=0$h) `MT' and fourth contact (`T4') points are marked in the colorbar of transit time. This illustrates that planetary material might be transiting before T1 (early-ingress) or after T4 (late-egress).
    The vertical red dashed lines show the three centers of the He I line.
    Models are identified in the top left corner.
    }
    \label{fig:Excess-abs}
\end{figure*}

We show in Figure \ref{fig:Excess-abs} the synthetic spectra of excess absorption (in percentage) as a function of wavelength for the He I line for four of our models.
The transmitted spectra are shown for different transit times, highlighting first contact (T1), mid-transit (MT) at $t=0$h and fourth contact (T4).
These times are marked in the colorbars of each plot, and in the figure, we display only the time where obscuration exists by the planetary material above an excess threshold of 0.05\%.

Models \swone\ and \swfour\ show the largest peak absorptions at \REV{$2.8\%$} and \REV{$3.8\%$}, respectively.
On the other hand, \swtwo\ and \swthree\ only reach \REV{$1.0\%$} and \REV{$0.8\%$} at mid-transit.
The weaker stellar wind model (\swone) shows clear pre-transit obscuration (early ingress), more than 3h prior to transit, with its T1 contact point having an excess absorption of around \REV{$1.1\%$}.
An early ingress is not seen for models of enhanced stellar wind (\swtwo\ and \swthree), as these have the bow shock quite close to the planet (i.e., the atmosphere does not extend substantially ahead of the orbital motion).
With the exception of models \swtwo\ and \swthree, all models presented in Table \ref{tab:models_props} show considerable obscuration after the last contact point T4, this is, in post-transit.
\swone\ shows a large blueshifted absorption of $\sim 0.3\%$ excess absorption up to $3\text{h}$ after T4.
Lastly, \swfour\ also exhibits absorption of \REV{$\sim 2.0\%$} prior to ingress and after egress.

Due to the 3D nature of our models, added by the geometric reshaping of the planetary material by interaction with stellar winds, we see an asymmetric distribution of outflowing material around the planet.
These bulk outflows generate spatially distinct regions of column densities with various line-of-sight velocities, ranging from $-120\, \text{km} \, \text{s}^{-1}$ to $+50\, \text{km} \, \text{s}^{-1}$, in the case of \swfour, for example.
In the case of a weaker stellar wind (\swone) the early ingress material (i.e., the frontward arm) is accelerated away from the planet towards the star, that is, away from the observer, causing the HeI absorption at T1 and prior ingress times to be mostly redshifted.
It is also possible to notice an enhanced redshift feature in model \swfour.
In contrast, at T4 and later egress times, only \swone\ shows a blueshifted absorption, emerging from most of the stream of gas in the tail being accelerated away from the planet.
This only happens when the nightside advected gas has a large enough column-density.
For cases of enhanced stellar wind (\swtwo\ and \swthree) or boosted stellar wind and XUV flux (\swfour), the tail is rarefied in triplet due to increase in temperature or stronger photoionization, respectively.

Moreover, models \swtwo\ and \swthree\ are also heavily influenced by interacting with stellar winds.
Both cases show absorption from strong blueshifted gas ($\sim -90\, \text{km} \, \text{s}^{-1}$), which originates an excess absorption of $\lesssim 0.1 \%$ at lower wavelengths ($\lambda < \lambda_1$).
Model \swtwo\ however, has a tail of material substantially distinct: the triplet density at the core \REV{can be up to an order} of magnitude higher compared to \swthree.
This allows for higher absorption around the HeI forming lines ($\lambda_{1,2,3}$), avoiding the dip between $\lambda_1$ and $\lambda_{2,3}$ that exists in model \swthree.
Adding to these, the temperature of the gas will cause thermal broadening.
As seen in Figure \ref{fig:2D-HDvars}, the temperature in the frontward arm of \swfour\ reaches $T\approx 10^4\text{K}$, which leads to a wider, although smaller, prior to mid-transit spectra.
%The nightside and behind-orbit outflow for both these models show an average smaller temperature, which impacts the profiles after T4, where the main feature comes from the doppler blueshift.
As the profiles in Figure \ref{fig:Excess-abs} are the summation of 3 lines, these plots contain multiple effects coming from a range of triplet densities, flow velocities and temperatures at each wavelength.

%We also investigated the variability of the peak absorption and EW for the models that do not reach a clear steady state.
%In the case of model \swone, we see the peak absorption during transit varying between $2.6\%$ and $2.8\%$, and EW ranging from $28$ to $31$ m\r{A}.
%This translates to a relative deviation from the average during transit for absorption or EW of less than $5\%$.
%The model \texttt{S1.6} sees the strongest changes, mainly due to the disruption of the extended bulk of planetary material up-orbit.
%Lastly, we see in \swfour that the peak varies between $1.55\%$ and $1.70\%$, or an EW ranging from $\sim 20$ to $22$ m\r{A}.
%A summary of the key properties of the synthetic spectra calculated for all models are shown in Table \ref{tab:model_summary}.

We also investigated the variability of the peak absorption and the \REV{equivalent width (EW)} for the models that do not reach a clear steady state.
%In the case of model \swone, we see the peak absorption during transit varying \REV{around $\pm 7\%$ over the time of evolution}, and EW ranging from \REV{$27$} to \REV{$32$} m\r{A}\REV{, translating to a relative deviation \REV{of $\sim 5\%$} from the average}.
%This translates to a relative deviation from the average during transit for absorption or EW of less than $5\%$.
In the case of model \swone, we see the peak absorption during transit varying \REV{around $\pm 7\%$ over the time of evolution}, and EW ranging \REV{within a relative deviation of $\sim 5\%$} from the average.
The model \texttt{S1.6} sees the strongest changes, mainly due to the disruption of the extended bulk of planetary material up-orbit.
%%%Lastly, we see in \swfour that the peak varies between $1.55\%$ and $1.70\%$, or an EW ranging from $\sim 20$ to $22$ m\r{A}.
Lastly, we see in \swfour\, that the peak \REV{EW vary less than $\pm 10 \%$}.
A summary of the key properties of the synthetic spectra calculated for all models are shown in Table \ref{tab:model_summary}.

\begin{table}
    \centering
    \REVcaption{Key properties of the synthetic transmission spectra: duration of obscuration by atmospheric material (which extends beyond T1 and T4), the minimum and maximum peak absorption between T1 and T4, and its time average, and the time average of EW for all models.}
    \begin{tabular}{ ||l||c||ccc||c|| }
        \hline
        Model & Duration & \multicolumn{3}{c}{Peak (\%)}  & EW (m\r{A}) \\
        & (h) & $ \min_\text{T} $ & $ \max_\text{T}$ & $<>_\text{T}$ & $<>_\text{T}$  \\ 
        \hline
        \hline
        %\hline
        \swone\ & $12.2$ & $1.1$ & $2.8$ & $2.5$  & $26.7$ \\ 
        %\hline  
        $\texttt{S1.6}$ & $11.6$ & $0.7$ & $1.9$ & $1.7$ & $20.4$ \\ 
        %\hline  
        $\texttt{S8.0}$ & $9.5$ & $0.3$ & $1.2$ & $1.1$ & $11.3$ \\ 
        %\hline  
        \swtwo\ & $4.9$ & $0.3$ & $1.0$ & $1.0$ & $8.1$ \\
        %\hline
        \swthree\ & $4.5$ & $0.3$ & $0.9$ & $0.8$ & $7.1$ \\
        \hline
        \swfour\ & $9.7$ & $2.1$ & $3.8$ & $3.3$ & $46.8$ \\
        \hline
        \hline
    \end{tabular}
    \label{tab:model_summary}
\end{table}

\begin{figure}
    \centering
        \includegraphics[width=\linewidth]{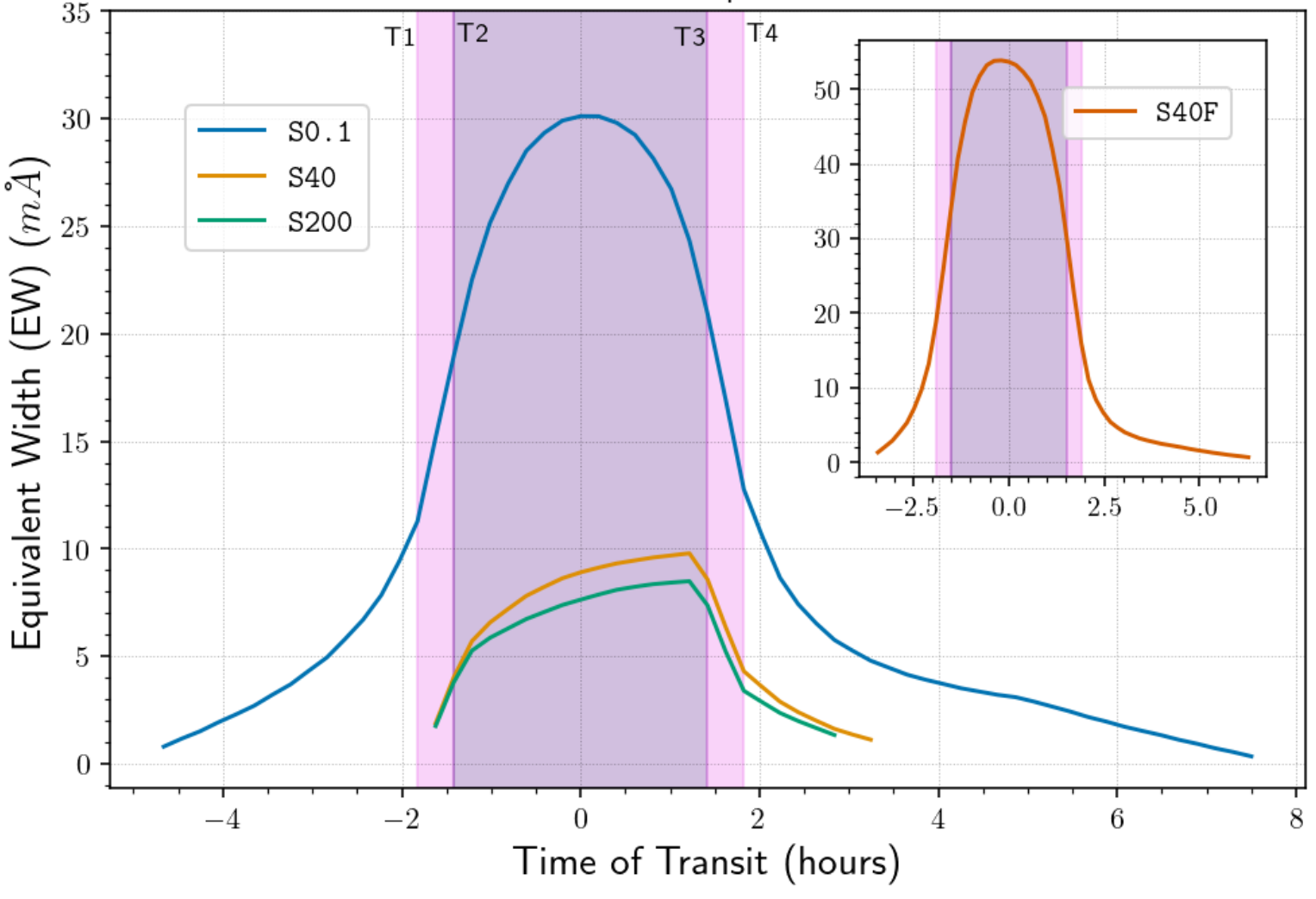}
    \REVcaption{
    Equivalent width (EW) of excess absorption of the He I line as a function of transit time for all models.
    The contact points of ingress (T1-T2) and egress (T3-T4) are annotated and shaded in light violet, and the T2-T3 transit in darker violet.
    }
    \label{fig:Synthetic-Compare}
\end{figure}

Figure \ref{fig:Synthetic-Compare} shows the EW of excess absorption as a function of time.
In this plot, vertical shaded regions show the contact points between T1 and T4.
We use an inset plot for model \swfour\ because it is a different stellar system, thus having different transit times.
As seen previously, \swone\ and \swfour\ have clear pre-transit signal, which indicates the presence of escaping material ahead of orbit.
We find a range of post-transit signals: when the volume of the tail is small, as seen for cases \swtwo\ and \swthree, EW quickly drops after transit.
Likewise, when the radiation environment is strong, the tail  reaches stronger ionization states, leading to a lower column-density.
From comparison between models, we confirm that comet-tail like outflows that have colder and less ionized gas provide for more extended light curves after egress.

\subsection{How do these results relate to observations?}
\label{SyntheticS_discussion}

\REV{Several transiting exoplanets have been observed with the HeI line in a variety of ground-based or space facilities.
Here, we bridge our synthetic results with an overall association to observed spectra.
}

\REV{Model \swone\, relates to systems with extended long tidal arms and tails, either due to a weak stellar wind or a strong tidal deformation, such as HAT-P-32b \citep[][]{23_Zhang---HAT-P-32b}, HAT-P-67b \citep[][]{24_Gully-Santiago---HAT-P-67b} and WASP-121b \citep[][]{24_Czesla---WASP-121b, 25b_Allart}.
Such systems tied to small Roche lobes are possible strongly evaporating exoplanets, which are expected to relate to several hours of excess absorption.
Additionally, strong photoevaporation ($\dot{\mathcal{M}}_p \gtrsim 10^{12} \text{g}/\text{s}$) can also relate to high stellar XUV flux, as seen by the $~20\times$ increase in mass loss from models \swtwo\ and \swfour.
Asymmetric transits, like case HAT-P-67b, can be a good probe to study the combination of tidal effects with stellar radiation, since the leading material obscures strongly for several hours in pre-transit, while the tail seems triplet deficient.
}

\REV{The stellar wind influence throughout models \texttt{S8.0}, \swtwo\, and \swthree\, is key to correctly characterize a planet's evaporation, or even to interpret some tentative detections.
From several dedicated programs at observing HeI transits, such as MOPYS \citep[][]{24_Orell}, GAPS \citep[][]{24_Guilluy}, and others \citep[][]{22_Vissapragada, 23_Allart, 26_Saidel}, it seems that commonly confirmed He evaporating targets are hot-/warm-/sub Neptunes, often with transit signals with small EWs and short periods of obscuration after their transit.
These can be interpreted as signature of comet-like tails, which are shaped due to the interaction of the planetary outflow with the stellar wind.
A compatible example with our simulations is the small hot-Jupiter WASP-69b \citep[see works by][]{18_Nortmann, 24_Tyler, 25_Allart}, that has a confirmed trailing tail of evaporating material, of which He is included.
Systems like these could be useful for retrieving stellar host mass-loss rate.
However, a more detailed evaluation seems necessary for interpreting data variability (e.g., whether due to variation in XUV flux, stellar wind, magnetic cycles, etc, \citealt{23_Krolikowski, 25_Sanz-Forcada, 25b_Allan}).
}

\REV{Lastly, from our models \swtwo, \swthree\, and \swfour, we found that a reduced density in He triplet can be due to strong wind confinement or by high mid-UV and X-ray radiation. 
This raises the question of what are the factors that promote an extremely small (or non-existent) transit observed in the HeI line.
Many such cases have been reported in the literature, of which some examples are: GJ 1214b, HD 97658b and GJ 9827d \citep[][]{20_Kasper}; TRAPPIST-1b, e, and f \citep[][]{21_Krishnamurthy}; K2-100b and V1298 Tau c \citep[][]{24_Alam} and HD 63433b with \citep[][]{22_Zhang, 24_Alam}, among others.
Non-detections of evaporating helium triplet are usually interpreted as lack of atmospheric escape or low helium abundance (e.g., secondary atmosphere).
For this reason, for cases of young planets, which are thought to possess helium-rich primary atmospheres,  
%When looking at young planets with primary atmospheres, 
the lack of a clear detection should be evaluated under the aforementioned effects \citep[see e.g.,][]{22_Fossati, 23_Fossati, 25_Allan}.
}

\section{Discussion}
\label{section6}

% - effects of stellar wind
%
% - helium chemistry: strongest impacts
% 
% - improvements and limitations

%*********************************************

\subsection{Stellar wind effects through the HeI line}

The stellar wind is paramount in shaping the transmission spectroscopy of atmospheric escape, as seen in Sections \ref{section4} and \ref{section5}.
%It is the pressure balance between the cold and dense planetary material with the hot and tenuous stellar wind that dictates the geometry of the outflow.
%Simultaneously, the chemical network introduced in this work is highly dependent on temperature and density of species.
%This leads to a population re-structuring in the shock zone, which we found is a region of helium triplet enhancement.
%Likewise, the region of mainly cold planetary gas holds most of the triplet column density.
\af{
Our 3D self-consistent model is simultaneously balancing pressure from the encounter of cold planetary material and the hot and tenuous stellar wind, with a chemical network highly dependent on temperature and density of species.
We now briefly outline the results of the previous two sections.
}

Firstly, we address how the stellar wind impacts the helium population.
We focus on models \swone, \swtwo\ and \swthree, as these only vary the stellar wind profile.
%The volumetric heating rate shows a consistent pattern among models: as far as the planetary outflow extends, recombination photons contribute the most for total heating power.
%Yet, volumetric cooling rate increases with increasing stellar wind strength, as illustrated in Figure \ref{fig:2D-energy}.
%As stellar winds induce compression to the volume of planetary material, PdV work triggers temperature gradients and, at the same time, photons become subject to different optical depths.
\af{
As stellar winds induce compression to the volume of planetary material, PdV work triggers temperature gradients. 
At the same time, photons become subject to different optical depths, which consequently impacts heating and photoionization.
}
%Thus, stellar winds have a strong impact on the outflow's overall temperature.
Consequently, different temperature profiles introduce specific He-population networks, affecting the fraction of $\text{He}(2^3S)$ to total He.
\af{
When stellar winds shock at considerable large distances from the planet, like the extended outflow of case \swone, we find a mainly cold ($T \lesssim 10^4 \text{K}$) outflow due to a large column density of neutral species and dominant planetary ram pressure.
}
The triplet's net rate is mainly dictated by recombination of $\text{He}^+$ to $\text{He}(2^3S)$ ($\alpha_{\text{B}}[\text{He}(2^3S)]$) and the collisional de-excitation of $\text{He}(2^3S)$ to the intermediate state of $\text{He}(2^1S)$ ($\xi[\text{He}(2^1S)]$).
%Consequently, since the recombination of $\text{He}^+$ to $\text{He}(2^3S)$ follows an inverse power-law for temperature, the model reproduces the expected large fraction of triplet.
%In the shock zone, the leading source and sink for the triplet are the collisional excitation of $\text{He}(1^1S)$ to $\text{He}(2^3S)$ ($\Xi[\text{He}(2^3S)]$) and the collisional ionization of $\text{He}(2^3S)$ to $\text{He}^+$ ($\Psi[\text{He}(2^3S)]$),  respectively.
%These rates do not play a major role for the isolated outflow, but become important when interacting with stellar winds.
\swtwo\ and \swthree\ models follow similar hierarchy of rates, although these rates may change within one order of magnitude as the stellar wind strength increases.
%We see these rates increasing at the shock zone, reaching rate values ($\text{s}^{-1}$) higher than values at the planetary boundary.
When the flow-flow boundary is at a sub-sonic distance, the gas subject to advection experiences smaller optical depths and higher temperatures, alongside higher electron number density.
This results in a comet-like outflow mainly composed of ionized He particles - as the singlet is likely to be photoionized, and the triplet is scarce due to a lack of cold $\text{He}^+$.
%A $T\approx2\times10^4 \text{K}$ dayside gas temperature (as in \swthree) will result in a comet-like outflow mainly composed of ionized He particles - as the singlet is likely to be photoionized and the triplet is scarce due to a lack of cold $\text{He}^+$.
So, hot and ionized stellar winds (i.e., winds of $\text{He}^{++}$ and $\text{H}^{+}$) contribute to enhance the amount of $\text{He}^+$, which leads the enhancement of triplet species once the gas has cooled down sufficiently.

% I droped this  part of comparing with 1D works, it just breaks the readibility of the text.

%When comparing model \swone\ with 1D works, we find similar population fractions (see figure 6 of \citealt{23_Allan} and Figure \ref{fig:1D-fraction-species} of the present work).
%However, our models do not retain the same volumetric rate profiles because we are using different rates for the collisional excitation rates, as well as a different population-schematic.
%- also discussed by \citet{22_Yan} and \citet{Biassoni--2023} when looking at the importance of assuming non-isothermal rates.
%This is justified by the usage of different rates for the collisional excitation rates - a matter already discussed by \citet{22_Yan} and \citet{Biassoni--2023} when looking at the importance of assuming non-isothermal rates.
%So, although the insertion of $\text{He}^{++}$ through the stellar wind can indeed contribute to enhance the amount of $\text{He}^+$, this will only populate the triplet species if the flow is cooled down sufficiently.
%- thus the contrasting difference in triplet relative density in model \swthree\ compared to \swtwo.
%As a result, the hot ($T=10^6$K) ionized stellar wind made up of $\text{He}^{++}$ contributes to enhance the amount of $\text{He}^+$, which will populate the triplet species once the gas has cooled down sufficiently.
%Due to the temperature power-law profile of $\alpha_{\text{B}}[\text{He}(2^3S)]$, the formation of triplet increases as the temperature decreases.

\af{
Secondly, the effects of different stellar XUV fluxes are clearly seen when we compare models \swtwo\ and \swfour, where the stellar wind profile is the same.
}
The model with a younger star (\swfour) reveals photoionization to be the leading sink of triplet after $r\gtrsim2.5\, R_p$, keeping its balance with recombination as the main source rate.
%In regard to the energy equation, the high XUV flux of model \swfour\ gives rise to a different regime from \swtwo, now characterized by volumetric heating rates dominated by photoionization with negligible influence from recombination photons.
\af{
We also note a rise in volumetric heating rates, dominated by photoionization and with negligible influence from recombination photons, when compared to \swtwo.
}
This boosts the mass-loss rate, allowing the volume of outflowing gas to grow and reach pressure balance at further distances.
Therefore, an increase in the XUV environment makes the planetary gas hotter and more ionized at closer distances.
\af{
We also see in model \swfour\  a comparable lack of helium triplet, due to mid-UV photoionization.
}

For pre-transit times (before T1), models \swone\ and \swfour\ show a redshifted signal, originated by the extended photoevaporated material.
%Both cases have ingress contributions mostly from $\lambda_{2,3}$ lines, while $\lambda_1$ line is only enhanced during the transit, since it probes deeper into the outflow.
%Case \swfour\ spectra presents itself thermally more broadened, although the excess obscuration is stronger in case \swone\ because of a larger column density of triplet.
\af{
Due to its flux-enhanced outflow, the spectra of case \swfour\ is thermally more broadened, but experiences a smaller column density.
This is expressed in the strong redshift feature above $\lambda = 10830$ \r{A}, when comparing mid-transit spectra of \swone\ with \swfour\ (Figure \ref{fig:Excess-abs}).
}
%Comparing the mid-transit (MT) spectra, the strong Doppler redshift component of \swfour\ (above $\lambda = 10830$ \r{A}) comes from the enhanced outflow directed towards the host star.
%
%When the planetary outflow is mainly shaped into a comet-like tail by the stellar wind - models \swtwo\ and \swthree\ - spectra have smaller equivalent widths.
%The absorption gap between $\lambda_{1}$ and $\lambda_{2,3}$ is bridged during transit for case \swtwo, since there is still a substantial fraction of triplet in the photoevaporated gas, while the opposite happens in \swthree.
%Thus, small line-of-sight velocities ($|v_\text{los}| \sim 10\, \text{km} \, \text{s}^{-1}$ require a relatively large column density in order to widen the line's obscuration.
\af{
When the planetary outflow is mainly shaped into a comet-like tail by the stellar wind - models \swtwo\ and \swthree\ - we expect spectra with smaller equivalent widths.
This is a consequence of small line-of-sight velocities with narrow column densities of triplet.
As such, we see in the spectra of  \swtwo\ (bottom left plot in Figure \ref{fig:Excess-abs}) that the summed contribution of Doppler and thermal broadening at $\lambda_{2,3}$ lines give rise to an egress spectra (T4) with a blueshift towards $\lambda_{1}$.
On the other hand, \swthree\ only sees obscuration centered around each line.
There is also a very shallow but relatively strong blueshift component (bellow $\lambda = 10829$ \r{A}) after egress, mainly due to hot and strongly advected gas further down in the tail.
}

In summary, broadening behaviors will be expressed differently through these $\lambda_{1, 2, 3}$ wavelengths, contributing to specific traits in the final line spectra.
An increase in mass-loss rate of the stellar host modifies the overall $\text{He}(2^3S)$ state by converging to a hotter and less populated outflow, which consequently gives rise to shallower and smaller spectra.
When comparing models \swtwo\ and \swfour, we are also able to extrapolate that the XUV environment is key to contrast with stellar winds: stronger photoionization increases the outflow's mass-loss but also rarefies the triplet state.

%Even though the excess absorption spectra for the \swfour\ model is high, it is not stronger because of the mid-UV flux that quickly ionizes the triplet, thus decreasing the effective column density of this neutral species over the vast volume used by the gas.
%The best identifiable characteristic that distinguishes \swfour\ from \swone\ is the stronger redshift, an enhanced feature of the line $\lambda_1 = 10829.09114$ \r{A} - which is associated to probing features closer to the planet - and a more time-abrupt EW profile.
%However, if obscuration only starts at the expected transit time and lasts for less than 4h in post-transit, then the stellar wind is likely to be mild or strong and the exoplanet atmospheric escape should not reach large frontward distances - in reference to the star.
%The only way to distinguish or restrict this broad classification is by studying the time-varying behavior of the excess absorption or the EW of the He I line - see Table \ref{tab:model_summary} for a quantitative comparison.

%%

%%%%%%%%%%%%%%%%%%

The impact of varying metallicity, stellar flux, planetary outflow, etc. \citep[see ][]{21_a_Wang, 21_Wang} raises the degrees of freedom when trying to understand observations with 3D models.
%With enough statistical analysis on transit geometry, as developed by the work of \citet{25_Gascon}, one can potentially restrict the strength of stellar wind with upper and lower limits.
%It is mostlikely possible to distinguish different stellar hosts by means of a deep analysis on time-dependent observational data, as the HeI line is made up of 3 individual lines - one line $\lambda_1$ deviated from the other two $\lambda_{2,3}$.
%From the results seen in the K-dwarf set, a higher ratio is expected for excess absorption from $\lambda_1$ relative to the excess absorption between $\lambda_{2,3}$ when the column density is shallow but stranded in a comet-like shaped hot outflow.
We note that all models presented here only account for a planet with a fixed large ratio of He to H, and yet, model \swthree\ expresses a \REV{$0.8\%$} excess absorption for a very strong stellar wind interaction.
Thus, lower values of $[\text{He}/\text{H}]_p$ and harsh environments - either with strong stellar winds or high mid-UV and X-ray stellar flux - may be a plausible explanation for non-detections.
%Nonetheless, as discussed in Appendix \ref{Section_Advectionless}, it is imperative to use a self-consistent model, otherwise a post-process approach and an advection-less scheme will underestimate the triplet ratio.
%Lastly, an enhancement of the triplet is seen in the shock zone for the intermediate and strong stellar wind cases, hinting to a limit case where a portion of the signal may be solely due to this region of increased number of triplet - even though the strong stellar wind case still has a significant contribution from the core of the cometary-like tail of the outflow.

\subsection{Model Limitations and Future Improvements}
\label{sec:Discussion-FutureImprovement}

The underlying goal of this work is to present a new 3D HD model that contains a self-consistent population-chemistry scheme of atomic hydrogen and helium, and interpret the impact of stellar winds and XUV fluxes.
However, the separate simulations presented are relatively computationally expensive: \REV{under ‘Steady State’ evolution,} models \swone, \swtwo\ and \swthree\ take around 48h to 72h of computing time with 240 processors, or up to 9 days for the case of \swfour.
\REV{For 2 minutes in ‘Time Accurate’, it took each model around 32h with the same processing power.}
%This is a clear setback when compared with 1D models, which are able to have higher resolution throughout the grid and add more chemical complexity.
%Our models were evolved in a time-explicit scheme, which converges differently from a time-implicit one.
%Many works fix over all space the relative abundance of He to H, in contrast to this work, where it is only fixed at the base and boundary levels.
%Nevertheless, our models spam a variation of $\leq 10\%$ to this abundance, thus hardening the level of trust in our results.
Also, the helium triplet \REV{seemed sensitive to grid resolution changes when using ‘Steady State’ evolution}, sometimes accumulating in boundaries of higher resolution adjacent to cells of lower resolution.
%It requires the implementation of an implicit time step scheme, although it has been tried out - a mere 1h was expected to last more than 24h of computational time, something that is not feasible.
%Even though an implicit time step scheme would translate in a more realistic overall population in the fluid, as soon as a run is changed - for example, when we wish to continue evolving a stopped run - the initial numerical step uses an Euler scheme, followed by the scheme that is requested to implement: a second order time evolution.
%This impacts the overall evolved population, loosing the previous chemistry-accurate solution.
%It may be possible to notice a granulation in the helium triplet density - Figure \ref{fig:1D-fraction-species} - but due to the reaction rates present, only two possible paths populate this species, turning out to be spatially harder to find a smooth net reaction rate - Figure \ref{fig:1D-net-rates} - even though, when looking in absolute values, the net rates behave in similar decaying forms.
\REV{Using `Time Accurate', we were able to improve these grid artifacts, although in model \swone, we still see the shadow of the grid} in the helium triplet net rates, seen as the discontinuous \REV{rectangular shape, with edges at $y=\pm 7.5R_p$,} seen in Figure \ref{fig:2D-SWs-Net-triplet-rates}. %or in the triplet profile (Figure \ref{fig:1D-net-rates}).
%\REV{%This discontinuity is related to the plane-parallel approximation and grid resolutions, specifically the $\vec{x}$-directed cylinder (pink object in Figure \ref{fig:grid}).
%We also see a faint jitter around a structure that follows the orbital path of the planet, but given that this follows the subsonic region (see density plot of Figure \ref{fig:2D-HDvars} for \swone), this indicates that this is less related to resolution jumps, and instead related to the local physical conditions.}
Note that only two possible paths populate this species, compared to 6 sinking rates, while having values much smaller than most other species.
The density profile of $\text{He}(2^3S)$ is the smallest for most of the grid and evolves through rates related to the amount of $\text{He}^+$ or $\text{He}(1^1S)$, present at much higher densities.
%Thus, we may be seeing stability artifacts when checking for all species net reaction rates, even though, when looking in absolute values, these net rates behave in similar spatial decaying forms.
%\REV{We evaluated this condition by comparing the directly computed net rate of triplet $\mathcal{R}[\text{He}(2^3S)]$ to the indirect net rate of triplet that follows from equations (\ref{EQ--mass-conservation}) and (\ref{EQ--population}): $\mathcal{R}[\text{He}(2^3S)] = -\sum_{ \{ \forall \text{sp} \, \setminus \, \text{He}(2^3S) \} } \mathcal{R}[\text{sp}] $.
%Nearly all cells are comfortably within a relative error of $\leq 5\%$. 
%The few exceptions appear at some cells in the borders of the geometrically resolved regions, randomly scattered over space and time.
%Additionally, complementing the information of Figure \ref{fig:1D-net-rates}, we confirm that the sum of all net rates - which should be 0.0 - stands at least several orders of magnitude smaller than each individual species net rate.
\REV{Typical timescales of the strongest rates ($\sim 10^2 \, \text{s}^{-1}$) translate to about $1.2 \times 10^4$ recombinations for a 2 minute evolution.
%When a ‘Time Accurate’ evolution scheme is applied, there is an overall smoothening of the triplet population, but the hydrodynamics become much slower.
}

Here, we do not explicitly include the helium states $2^1S$ or $2^1P$, as done in \citet{23_Allan, 25_Schulik}, meaning that we assume an immediate transition rate of triplet to singlet, as done in other works in the literature \citep[e.g.,][]{18_Oklopcic, Lampon--2020, 22_DosSantos}.

Additionally, we also lack molecular elements such as $\text{H}_2$ or $\text{HeH}^+$, which could impact the regions of subsonic outflow with additional cooling contributions (\citealt{25_Munoz, 26_Taylor}).
This is a clear setback when compared with 1D models, which are able to have higher resolution throughout the grid, add more chemical complexity or even use multi-fluids instead of multi-species in a single fluid.
%In a qualitative comparison with 1D multi-fluid models that have studied the metastable triplet in atmospheric escape \citep[][]{19_Khodachenko, 23_Xing, 25_Schulik}, we see some overall similarities and differences.
For example, the work of \citet{23_Xing} is suggestive towards mass fractionation that can happen for low He to H ratio values - potentially explaining observations, also discussed in \citet{25_Schulik}.
Despite the fact that this cannot be retrieved by our model, from their outputs it is possible to state that our model is realistically viable when assuming large ratios of He compared to H.
%Perhaps the spatial region where this work diverges from a framework of multi-fluids can be the zone of interacting planetary and stellar material, whereas close to the planet the outflow may be different in nature, but should have similar behavior.

Albeit several works have shown that collisional excitation and de-excitation of the metastable state $\text{He}(2^3S)$ are meaningful sinks for the population of this species, photoionization is also important. % - as seen by the different sets of models.
The cross-section coefficients used here are derived from fits to previous theoretical computations.
The work of \citet{25_Taylor} overcomes low-resolution coefficients by adopting a method that computes high-resolution cross-section coefficients that are energy dependent.
As the authors of the latter study commented, this allows for the appearance of artifacts such as resonances at wavelengths ranges overlapping with stellar coronal emission lines.
%In the models developed here, however, we limit the photoionization of triplet to the hEUV wavelength, assuming negligible cross-sections for the X-ray range.
%This is supported by the idea that most photoionized triplet will be due to mid-UV photons, which have flux values usually 3 order of magnitude higher.
%{ \color{red}{cite} }
Thus, higher resolution in photoionization cross-sections can be relevant in a context of resolved stellar spectra (see, e.g., the impact of the stellar host on HeI transits in \citealt{25b_Allan}).
However, this is less relevant for our models, as we use 4 channels of wavelength for the stellar spectra. This simplification, nevertheless, allows us to identify the leading heating sources.

%Finally, the presence of magnetic fields - not included here - have can alter the atmospheric escape of the exoplanet, as well as the transmission spectra in hydrogen transits (\citealt{21_Carolan, 21_Khodachenko}).
Finally, the presence of magnetic fields - not included here - \REV{can alter} the atmospheric escape of the exoplanet, as well as the transmission spectra in hydrogen transits (\citealt{21_Carolan, 21_Khodachenko}).
Depending on the geometry of the magnetic field, a fraction of magnetic field lines will be open \citep[][]{11_Adams}.
It will be through these open lines that ionized material escapes, while on the other hand, a dead-zone is formed around the planet by closed field lines.
Given that in our currently built model, we have two additional ionized species ($\text{He}^+$ and $\text{He}^{++}$) that affect the triplet population, the overall column density of triplet should be modified by a new flow behavior. 
This will be investigated in a future work.
Several studies have also shown that there is an impact in the mass-loss rate of the hydrodynamic escape \citep[][]{14_Owen, 21_Carolan} when in a magnetized scenario.
The work of \citet{24_Presa} brings additional insight to the importance of magnetic fields: the planetary magnetic obliquity were shown to impact the overall geometry of the outflow, and the Alfvenic regime in which the exoplanet exists is of relevance for the bulk escape.

\section{Conclusion}
\label{section7}

%In this paper, we  developed a 3D radiation HD model that self-consistently solves for population balance of atomic species: hydrogen ($\text{H}_0$; $\text{H}^+$) and He (singlet: $\text{He}(1^1S)$; triplet: $\text{He}(2^3S)$; singly ionized: $\text{He}^+$; doubly ionized: $\text{He}^{++}$).
In this paper, we developed a \REV{3D HD model that incorporates the effect of radiation in heating and cooling processes} while self-consistently solving for population balance of atomic species: hydrogen ($\text{H}_0$; $\text{H}^+$) and He (singlet: $\text{He}(1^1S)$; triplet: $\text{He}(2^3S)$; singly ionized: $\text{He}^+$; doubly ionized: $\text{He}^{++}$).

We have conducted two sets of models, each set with a different stellar host but maintaining the same exoplanet - a tidally locked Hot Jupiter located at $a=0.05\text{au}$ - with the goal of exploring the impact of stellar winds and XUV flux on the He I line.
The first set of models have a moderate XUV spectra and the exoplanet is subject to stellar winds of different mass-loss rates, while the second set assumes a younger star, with much higher XUV fluxes, embedded in a  wind that has a mass-loss rate that is $10$ times larger than the solar wind.
From the hydrodynamic outputs of each simulation, we are able to evaluate the significant impact that stellar winds have on the morphology of the outflow and specially, the influence on chemistry relevant to the He species.

This study revealed that the population of triplet is affected by stellar winds, \aav{with a decrease of helium triplet density with increasing stellar wind strength.}
%\af{
%The outflow bulk volume emerges from the pressure balance of planetary gas and stellar wind. 
%Accordingly, the amount of neutral material that resides inside this volume sets the optical depth of the gas, which impacts photoionization and volumetric heating rates.
%The pressure balance will impact optical depths, which affect volumetric heating.
%Importantly, the gas temperature is established by the interplay of volumetric heating and cooling rates with the expansion or contraction work on the gas.
%}
%Furthermore, expansion or contraction work on the gas will set a gas temperature that will balance with cooling (heavily temperature dependent components) and heating.
%The converged local temperature value then dictates the helium species arrangement, along with electron and depopulated species densities.
For a weak stellar wind ($1/40 \times$ solar), the relative fraction of triplet at the core of the escaping atmosphere  can be $n_{\text{He}(2^3S)} / n_\text{He} \sim 1 \times 10^{-5}$, while in a scenario of strong stellar wind ($50 \times$ solar) we found \REV{$n_{\text{He}(2^3S)} / n_\text{He} \sim 8 \times 10^{-7}$}.
\aav{This is because, as a strong stellar wind confines more the extension of the atmosphere, the optical depth for ionizing stellar photons is reduced and planetary material gets hotter, reducing the population of neutral and triplet helium.}
% as the triplet is only populated by $\text{He}^+$ recombination (at relatively low temperatures) and collisional de-excitation of singlet into triplet by means of a free electron.
\aav{
On the other hand, with a constant stellar wind mass-loss rate, but increasing the XUV flux, we showed that the escape rate of the outflowing material increases, as expected, along with an increase in helium absorption. 
% the gas is subject to changes in the volumetric heating rate and photoevaporation rates, reorganizing the ionization profiles in the outflow.
% This triggers an adjustment in the mass-loss rate, which, in the case of similar stellar winds but enhanced flux, evokes an increased excess in He I transit.
}
In the end, the stellar wind strength impacts the mass-loss rate of the planet and uniquely affects the chemistry network at hand.
\aav{We also showed that advection plays an important role in the triplet population, for example, in the nightside, as it brings $\text{He}^+$ from elsewhere to this region. For the triplet population, we found that neglecting advection leads to an over-estimation of triplet closer to the planet ($\lesssim 2\, R_p$),  and an under-estimation beyond $\gtrsim 2\, R_p$ (Appendix \ref{Section_Advectionless}.)}

Additionally, we  used a ray tracing model to compute the excess absorption of the transit of the exoplanet for the HeI line.
%With these models, we compared the synthetic spectra of our simulations to interpret the HD effects under the observational component of HeI line for atmospheric escape.
The distinct geometry of the interaction with stellar winds and the column density of the triplet species resulted in different excess absorption profiles that show quite distinct features over the transit duration.
Models with material orbiting ahead of the planet show pre-transit signal, and there is always non-negligible post-transit signal, because of a trailing comet-like tail.
During transit, according to each model, we can have blue and redshift features broadened by thermal motions and line-of-sight velocities from the wind-wind interaction.
A clear distinction is the model of strongest stellar wind, where excess absorption is less than $1\%$ and the three lines do not blend in wavelengths between them.
Potentially, for lower ratios of $\text{He}/\text{H}$, such winds may present as a valid cause to justify non-detections.
%In the two sets of models we encountered some similarities in the average transit spectra, which can be attributed to the allowed level of degeneracy by introducing stellar winds: a high column density in outflows that exceed the Hill radius by a factor of 4 can produce signals apparently similar as a lower column density to a quite extended outflow.
%Lastly, an enhancement of the triplet is seen in the shock zone for the intermediate and strong stellar wind cases, hinting to a limit case where a portion of the signal may be solely due to this region of increased number of triplet - even though the strong stellar wind case still has a significant contribution from the core of the cometary-like tail of the outflow.

%%%%%%%%%%%%%%%%%%%%%%%%%%%%%%%%%%%%%%%%%%%%%%%%%%
%%%%%%%%%%%%%%%%%%%%%%%%%%%%%%%%%%%%%%%%%%%%%%%%%%
%%%%%%%%%%%%%%%%%%%%%%%%%%%%%%%%%%%%%%%%%%%%%%%%%%
%%%%%%%%%%%%%%%%%%%%%%%%%%%%%%%%%%%%%%%%%%%%%%%%%%

\section*{Acknowledgements}

\REV{We thank the reviewer for their valuable comments and suggestions, which helped improve our paper.}
The authors thank Dr Andrew Allan for his suggestions and discussions over the course of this work. This publication is part of the project "Blowing in the wind: exoplanetary atmospheres being carried away by stellar winds" (with project number VI.C.232.041 of the research programme "NWO Talent Programme VICI 2023"), which is financed by the Dutch Research Council (NWO). AAV acknowledges funding from the European Research Council (ERC) under the European Union's Horizon 2020 research and innovation programme (grant agreement no. 817540, ASTROFLOW). This work used the Dutch national e-infrastructure with the support of the SURF Cooperative using grants nos. EINF-7488 and EINF-13632. This work used the BATS-R-US tools developed at the University of Michigan Center for Space Environment Modeling and made available through the NASA Community Coordinated Modeling Center.

%%%%%%%%%%%%%%%%%%%%%%%%%%%%%%%%%%%%%%%%%%%%%%%%%%
\section*{Data Availability}

The data used in this article will be shared on reasonable request to the corresponding author.
\af{
Animations of the performed simulations can be found in the following \href{https://doi.org/10.5281/zenodo.21455405}{Zenodo link}.
}



\bibliographystyle{mnras}
\bibliography{example} % if your bibtex file is called example.bib

%%%%%%%%%%%%%%%%%%%%%%%%%%%%%%%%%%%%%%%%%%%%%%%%%%

%%%%%%%%%%%%%%%%% APPENDICES %%%%%%%%%%%%%%%%%%%%%

\appendix

\section{Tables with relevant reactions}

\begin{table*}
    \centering
    \caption{The rates of populating and depopulating processes for the hydrogen and helium states considered in our model. }
    \label{tab:rates}
    \renewcommand{\arraystretch}{1.75} % Increases row spacing for better readability
    \setlength{\tabcolsep}{11pt} % Adjust horizontal spacing between columns
    \begin{tabular}{ p{1.0cm} | p{1.75cm} p{1.75cm} p{10cm} }
        \hline
        \textbf{Refs.} & \textbf{Populated Species} & \textbf{Depopulated Species} & \textbf{ \; \; \; Rates ($\text{g}\, \text{cm}^{-3} \, \text{s}^{-1}$)} \\
        \hline
        \hline

        \multicolumn{4}{c}{\textit{Recombination $\mathbf{\alpha}$ } } 
        \\
        a, b    &    $\text{H}_0$      &    $\text{H}^+$      &  
        $
        \alpha_\text{B}  [\text{H}_0] = 2.59 \times 10^{-13}  \left( \frac{T}{10^4} \right)^{-0.7}\, \text{n}_\text{e}\, \rho_{\text{H}^+}
        $
        \\
        c & He(1$^1$S)        & He$^+$          &  
        $
        \alpha_\text{B} [\text{He}(1^1S)] = 6.23 \times 10^{-14} \left(\frac{T}{10^4}\right)^{-0.827}\, \text{n}_\text{e}\, \rho_{\text{He}^+}
        $
        \\
        c & He(1$^1$S)        & He$^+$          & 
        $
        \alpha_\text{A-B}$ [He(1$^1$S)] = $1.54 \times 10^{-13} \left(\frac{T}{10^4}\right)^{-0.486}\, \text{n}_\text{e}\, \rho_{\text{He}^+}
        $  \par (emits 24.59 eV photon) 
        \\
        c & He(2$^3$S)        & He$^+$          & 
        $
        \alpha_\text{B}$ [He(2$^3$S)] = $2.10 \times 10^{-13} \left(\frac{T}{10^4}\right)^{-0.778}\, \text{n}_\text{e} \, \rho_{\text{He}^+}
        $    
        \\
        d, e  & He$^+$             & He$^{++}$                & 
        $
        \alpha_\text{B}$ [He$^+$] = $5.506 \times 10^{-14} \left(\frac{1263030}{T}\right)^{1.5} \left(1+\left(\frac{460960}{T}\right)^{0.407} \right)^{1.5} \, \text{n}_\text{e} \, \rho_{\text{He}^{++}}
        $    
        \\
        \hline
        \multicolumn{4}{c}{\textit{Collisional ionization  $\mathbf{\Psi}$ }} 
        \\
        f, g & $\text{H}^+$             & $\text{H}_0$                & 
        $
        \Psi$ [$\text{H}_0$] = $ 1.27 \times 10^{-21} \sqrt{T} \exp[-\frac{157809.1}{T}] \,\text{n}_\text{e} \,\rho_{\text{H}_0} \big{/} \left( \text{E}_{\text{H}_0} \right) 
        $     
        \\
        f, g & He$^+$             & He($1^1$S)                & 
        $
        \Psi$ [He($1^1$S)] = $ 9.38 \times 10^{-22} \sqrt{T} \exp[-\frac{285335.4}{T}] \, \text{n}_\text{e} \, \rho_{\text{He}(1^1\text{S}) } \, \big{/} \left( \text{E}_{\text{He}(1^1\text{S})} \right) $ 
        \\
        f, g & He$^+$             & H($2^3$S)                & 
        $
        \Psi$ [He($2^3$S)] = $ 6.41 \times 10^{-21} \sqrt{T} \exp[-\frac{55338}{T}] \, \text{n}_\text{e} \, \rho_{\text{He}(2^3\text{S})} \, \big{/} \left(\text{E}_{\text{He}(2^3\text{S})} \right) 
        $     
        \\
        f, g & He$^{++}$             & He$^+$                & 
        $
        \Psi$ [He$^+$] = $ 4.95 \times 10^{-22} \sqrt{T} \exp[-\frac{631515}{T}] \, \text{n}_\text{e} \, \rho_{\text{He}^+} \, \big{/} \left( \text{E}_{\text{He}^+} \right) 
        $     
        \\
        \hline
        
        \multicolumn{4}{c}{\textit{Collisional de-excitation $\mathbf{\Xi}$ }} 
        \\
        h, i & He(1$^1$S) & He(2$^3$S)                & 
        $
        \Xi$ [He(1$^1$S)] = $5.0 \times 10^{-10}  \, \text{n}_{\text{H}_0} \,\rho_{\text{He}(2^3\text{S}) } 
        $  \par (by means of a $\text{H}_0$)
        \\ 
        e, h, i, j   & He(2$^3$S) & He(1$^1$S)                & 
        $
        \Xi$ [He(2$^3$S)] = $ 2.10 \times 10^{-8} \sqrt{ \frac{13.6}{\text{k}_\text{B} T} } \exp{\left( - \frac{19.81}{\text{k}_\text{B} T} \right) } \,\Upsilon_{13} \, \text{n}_\text{e} \,\rho_{\text{He}(1^1\text{S}) } 
        $ \par (by means of a e$^-$)       
        \\
        
        \hline
        
        \multicolumn{4}{c}{\textit{$2^1S$ and $2^1P$ states Collisional de-excitation $\mathbf{\xi}$ }} 
        \\
        e, i, j & He(2$^1$S) & He(2$^3$S)                & $\xi$ [He(2$^1$S)] = $2.10 \times 10^{-8} \sqrt{ \frac{13.6}{\text{k}_\text{B} T} } \exp{ \left( - \frac{0.80}{\text{k}_\text{B} T} \right)} \, \Upsilon_{31A} \, \text{n}_\text{e} \, \rho_{\text{He}(2^3\text{S}) } $ \par (by means of a e$^-$)       
        \\
        e, i, j & He(2$^1$P) & He(2$^3$S)                 & $\xi$ [He(2$^1$P)] = $2.10 \times 10^{-8} \sqrt{ \frac{13.6}{\text{k}_\text{B} T} } \exp{ \left( - \frac{1.40}{\text{k}_\text{B} T} \right) } \, \Upsilon_{31B} \, \text{n}_\text{e} \, \rho_{\text{He}(2^3\text{S}) } $ \par (by means of a e$^-$)       
        \\
        
        \hline
        
        \multicolumn{4}{c}{\textit{Charge exchange $\mathbf{\chi}$ }} 
        \\
        h, k &  $\text{H}^+$, He(1$^1$S)        & $\text{H}_0$, He$^+$         & $\chi$ [He(1$^1$S)] = $1.25 \times 10^{-15} \left(\frac{T}{300} \right)^{0.25} \,m_\text{H} \, \text{n}_{\text{H}_0} \, \text{n}_{\text{He}^+} $      
        \\
        h, k &  $\text{H}_0$, He$^+$        & $\text{H}^+$, He(1$^1$S)         & $\chi$ [He$^+$] = $1.75 \times 10^{-11} \left(\frac{T}{300} \right)^{-0.75} \exp[-\frac{128000}{T}] \, m_\text{H} \, n_{\text{H}^+} \, n_{\text{He}(1^1\text{S})} $      
        \\
        \hline
        \multicolumn{4}{c}{\textit{Radiative Decay $\mathbf{\varphi}$ }} 
        \\
        l &  He(1$^1$S) & He(2$^3$S)             & $\varphi$ [He(1$^1$S)] = $1.272 \times 10^{-4} \, \rho_{\text{He}(2^3\text{S}) } $      
        \\
        \hline
    \end{tabular}
    
    \vspace{0.5cm}
    \footnotesize
    \textbf{Notes:}  References: References: a -- \citet{Storey-Hummer--1995}, b -- \citet{Oster-Fer--2006}, c -- \citet{Benjamin--1999}, d -- \citet{Hui-Gnedin-1997}, e -- \citet{Caldiroli--2021}, f -- \citet{Black--1981}, g -- \citet{Cen--1992}, h -- \citet{Bray--2000}, i -- \citet{18_Oklopcic}, j -- \citet{Lampon--2020}, k -- \citet{Koskinen--2013}, l -- \citet{Drake--1971}.
    We emphasize that the behavior of $\Upsilon_{ij}$ is temperature dependent and is fitted by \cite{Caldiroli--2021} - presented explicitly in GitHub: \href{https://github.com/AndreaCaldiroli/ATES-Code/tree/main/src/modules/radiation}{Link}.
\end{table*}

Tables \ref{tab:rates} and \ref{tab:cooling_processes} provide the expression for the (de)populating rates and volumetric cooling rates used in our model. 
Note that many of the reaction rates present in Table \ref{tab:rates} should have temperature distinction between electron, ion or fluid temperatures. 
Due to the nature of our single-fluid model, we can only retrieve a fluid averaged temperature. 

\begin{table*}
    \centering
    \caption{Cooling processes included in our model in terms of volumetric cooling rate. The sum of the cooling rates ($\mathcal{C}$) is included in Equation (\ref{EQ--energy-conservation}). Number densities are given in cm$^{-3}$ and temperatures in K.}
    \label{tab:cooling_processes}
    
    \renewcommand{\arraystretch}{1.75}
    \setlength{\tabcolsep}{11pt}
    \begin{tabular}{| l |||||| p{10cm} }
        
        \hline
        
        \textbf{References} & \textbf{Volumetric cooling rates ($\text{erg}\, \text{cm}^{-3} \, \text{s}^{-1}$)}  \\
        
        \hline
        \hline
        
        \multicolumn{2}{c}{\textit{Collisional ionization cooling}} 
        \\
        a,b & $1.27 \times 10^{-21} \sqrt{T} \exp[-\frac{157809.1}{T}]  \, \text{n}_\text{e} \, n_{\text{H}_0}$
        \\ 
        a,b & $9.38 \times 10^{-22} \sqrt{T} \exp[-\frac{285335.4}{T}]  \, \text{n}_\text{e} \, n_{\text{He}(1^1\text{S})}$
        \\ 
        a,b & $6.41 \times 10^{-21} \sqrt{T} \exp[-\frac{55338}{T}] \, \text{n}_\text{e} \, n_{\text{He}(2^3\text{S})}$ 
        \\ 
        a,b & $4.95 \times 10^{-22} \sqrt{T} \exp[-\frac{631515}{T}] \, \text{n}_\text{e} \, n_{\text{He}^+}$ 
        \\ 
        \hline
        \multicolumn{2}{c}{\textit{Recombination cooling}} 
        \\
        b & $ 8.70 \times 10^{-27} \sqrt{T} \left( \frac{T}{10^3} \right)^{-0.2} \, \text{n}_\text{e} \,  n_{\text{H}^+}$ 
        \\ 
        a,b & $1.55 \times 10^{-26} T^{0.3647} \, \text{n}_\text{e} \, n_{\text{He}^+}$
        \\ 
        a,b & $ 3.48 \times 10^{-26} \sqrt{T} \left( \frac{T}{10^3} \right)^{-0.2}  \, \text{n}_\text{e} \, n_{\text{He}^{++}}$
        \\ 
        
        \hline
        \multicolumn{2}{c}{\textit{Dielectronic recombination cooling}} \\
        a,b & $1.24 \times 10^{-13} T^{-1.5} \exp[-\frac{470000}{T}] \left( 1 + 0.3 \exp[-\frac{94000}{T}] \right) \, \text{n}_\text{e} \, n_{\text{He}^+}$
        \\ 
        
        \hline
        \multicolumn{2}{c}{\textit{Collisional excitation cooling}} \\
        a,b,c,d & $7.5 \times 10^{-19} \exp[-\frac{118348}{T}] \, \text{n}_\text{e} \, n_{\text{H}_0}$ 
        \\ 
        a,b & $5.54 \times 10^{-17} T^{-0.397} \exp[-\frac{473638}{T}] \, \text{n}_\text{e} \, n_{\text{He}^+}$ 
        \\ 
        a,b & $1.16 \times 10^{-20} \sqrt{T} \exp[-\frac{13179}{T}] \, \text{n}_\text{e}\, n_{\text{He}(2^3\text{S})}$
        \\ 
        
        \hline
        \multicolumn{2}{c}{\textit{Free-free emission (Bremsstrahlung)}} \\
        a,b & $1.42 \times 10^{-27} \sqrt{T} \left( \frac{3}{2} \right) \, \text{n}_\text{e} [n_{\text{H}^+} + n_{\text{He}^+} + 4n_{\text{He}^{++}}] $
        \\ 
        
        \hline
        
    \end{tabular}
    
    \vspace{0.5cm}
    \footnotesize
    \textbf{Notes:}  References: a -- \citet{Black--1981}, b -- \citet{Cen--1992}, c -- \citet{MurrayC--2009}, d -- \citet{19_Allan}.
\end{table*}

\section{Numerical methods}
\label{appendix_exp-tau}

The treatment of the exponential of optical depths (Equation (\ref{EQ--optical-depth})) is done through the use of a normalization value. 
Note that the advantage of performing this treatment is that we only need to compute a total of 4 different integrations - one for each species - instead of an integration for each species at each channel.
%, as a way to better deal with numerical truncation.
Assuming a typical cross-section base value of $\sigma_0 = 10^{-18}\text{barn}$, we computed as follows:
%
%\begin{equation}
\begin{align*}
\exp{(-\tau_{\lambda})} & = \exp{\left(- \sum_{\text{sp}} \tau_{\lambda}[\text{sp}] \right)}
\\ & = \prod _\text{sp} \exp{ \left(- \int _{\text{start } x} ^{\text{final } x} n_\text{sp}  \sigma_{\text{sp}, \lambda} dx \right)} \\
& = \prod _\text{sp} \exp{ \left( - \frac{\sigma_{\text{sp}, \lambda}}{\sigma_{0}} \int _{\text{start } x} ^{\text{final } x}  \sigma_{0} n_\text{sp}   dx \right)} = \\
& = \prod _\text{sp} \left( \exp{ \left( - \int _{\text{start } x} ^{\text{final } x}  \sigma_{0} n_\text{sp} dx \right) }\right)^{\left( \frac{\sigma_{\text{sp}, \lambda}}{\sigma_{0}} \right) } 
\end{align*}
\label{EQ--exp(-tau)}
%\end{equation}

%\aav{it is still not clear to me why you need this treatment and what is the relevance of the 1e-18 value you quote before...}
% AnsJGFal: In order to avoid powers of floats very close to 0, I chose a sigma0 that helps in this manner.

\section{Additional figures}
\label{appendix_Extra-Figs}

%\aav{you need to describe here what each figure shown below means}

\begin{figure*}
    \centering
        \includegraphics[width=\linewidth]{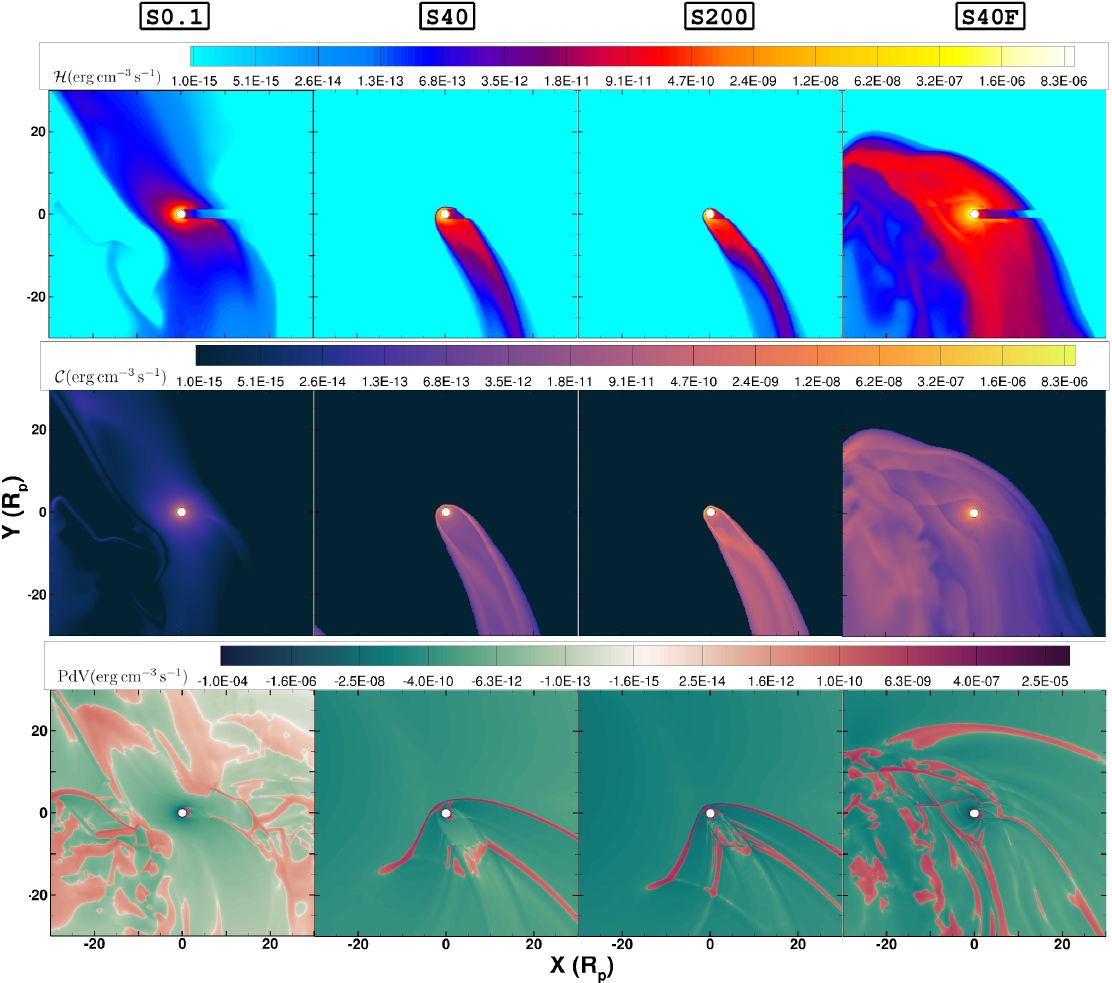}
    \caption{2D view in the orbital plane $z=0$ of the total volumetric heating ($\mathcal{H}$) and cooling ($\mathcal{C}$) rates, and PdV work in the top, middle and bottom row of panels, respectively.
    Each column corresponds to a different model. 
    }
    \label{fig:2D-energy}
\end{figure*}

Figure \ref{fig:2D-energy} displays the important components of the energy equation: total volumetric heating rate (top), total volumetric cooling rate (middle) and PdV work (bottom panels).

\begin{figure*}
    \centering
        \includegraphics[width=\linewidth]{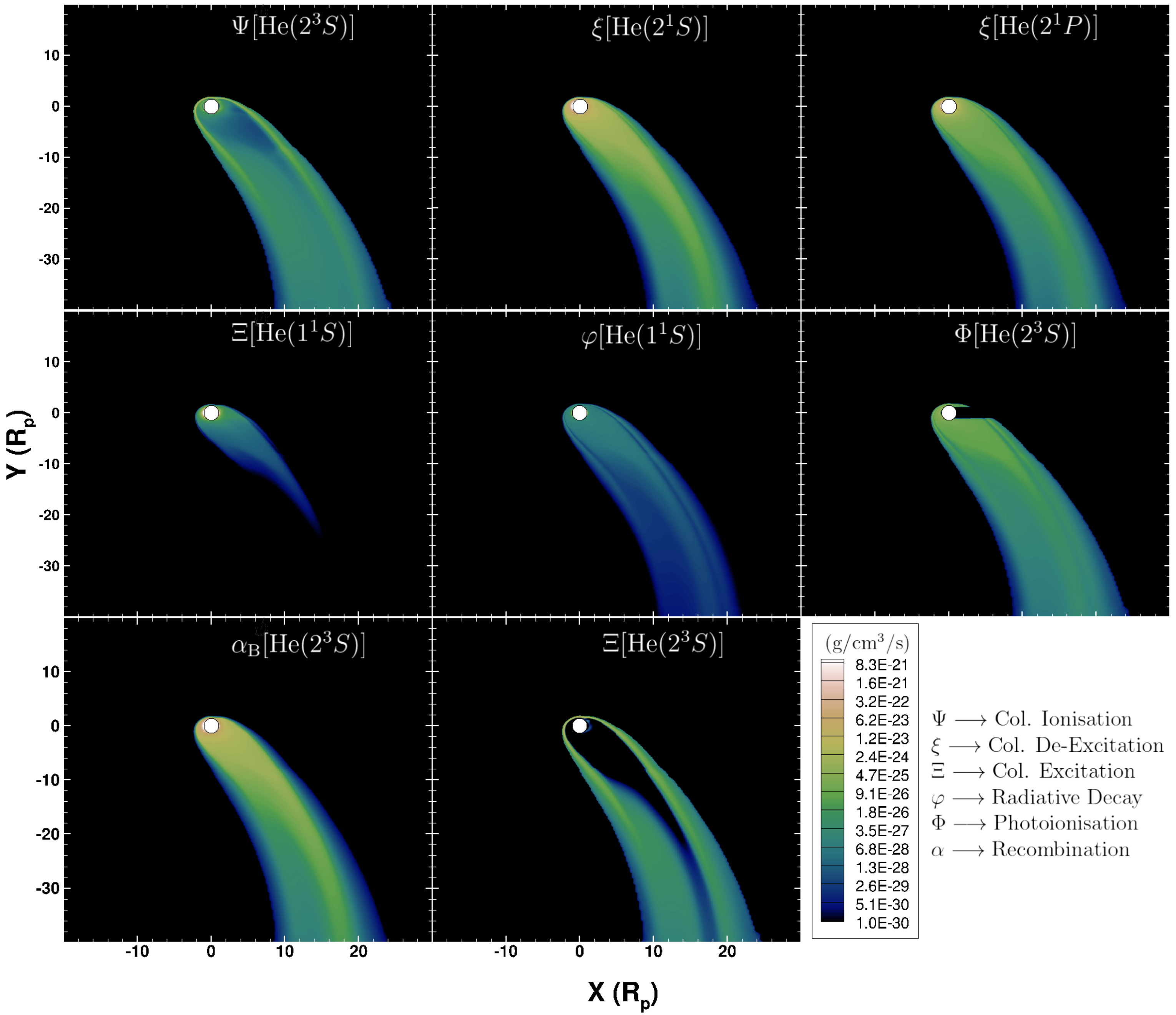}
    \REVcaption{2D view in the orbital plane $z=0$ of all contributing reactions rates to the population of He triplet for the model \swtwo, centered on the planet.
    Top and middle rows: annihilating rates.
    Bottom row: populating rates.
    The net rate can be found as the \REV{upper right} plot in Figure \ref{fig:2D-SWs-Net-triplet-rates}.}
    \label{fig:2D-SW2-All-triplet-rates}
\end{figure*}

\begin{figure*}
    \centering
        \includegraphics[width=\linewidth]{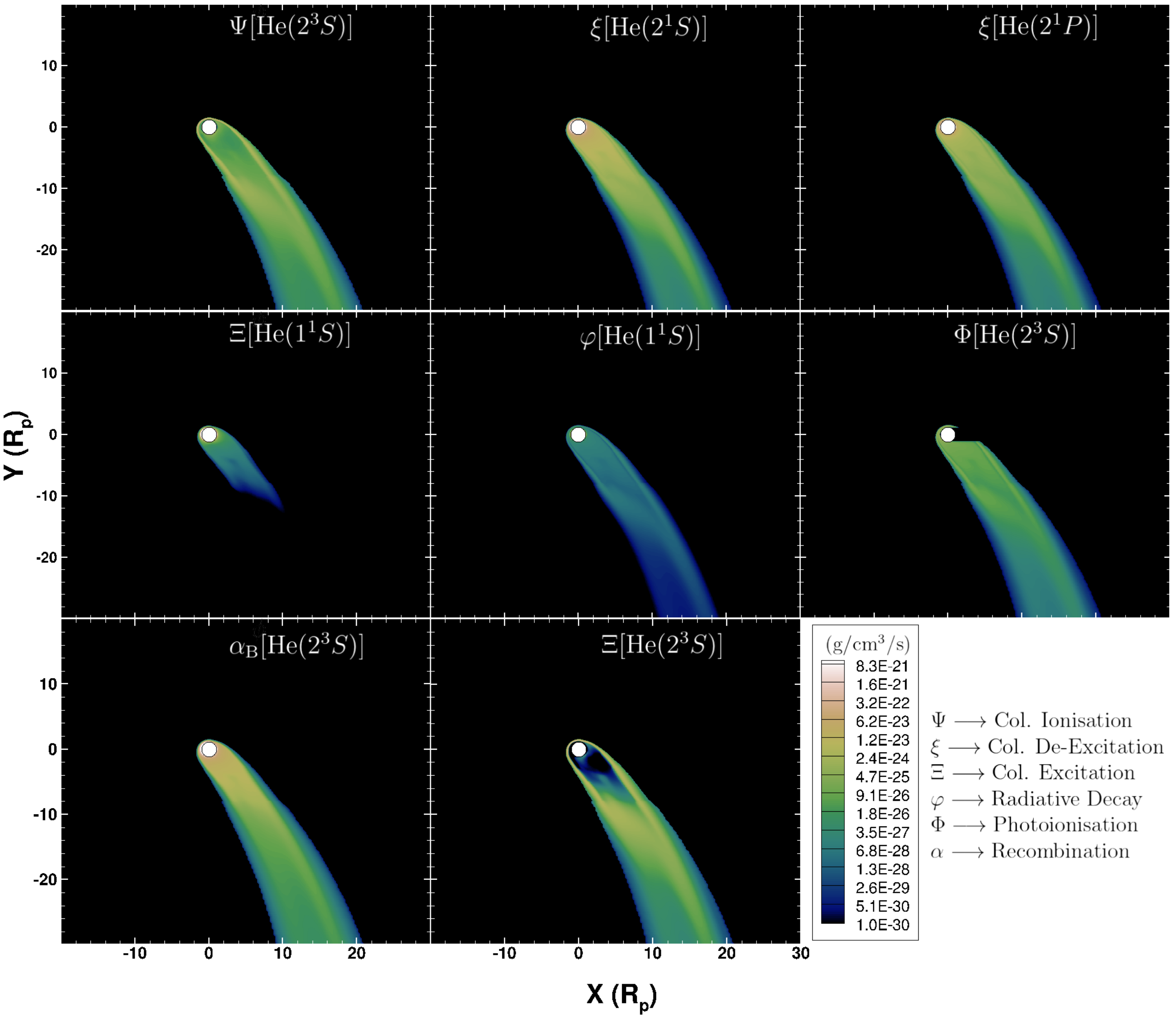}
    \REVcaption{2D view in the orbital plane $z=0$ of all contributing reactions rates to the population of He triplet for the model \swthree, centered on the planet.
    Top and middle rows: annihilating rates.
    Bottom row: populating rates.
    The net rate can be found as the \REV{lower left} plot in Figure \ref{fig:2D-SWs-Net-triplet-rates}.}
    \label{fig:2D-SW3-All-triplet-rates}
\end{figure*}

\begin{figure*}
    \centering
        \includegraphics[width=\linewidth]{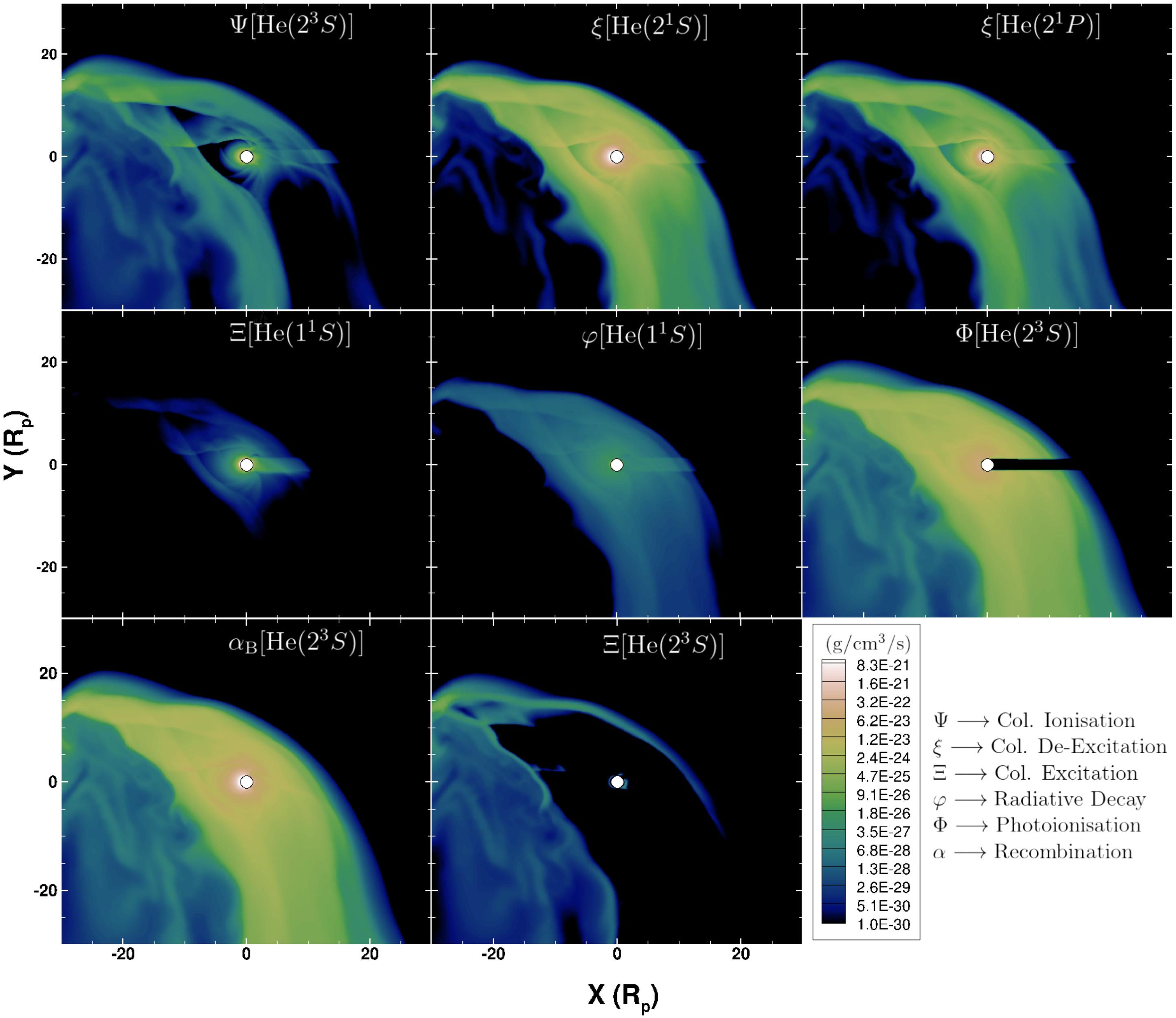}
    \REVcaption{2D view in the orbital plane $z=0$ of all contributing reactions rates to the population of He triplet for the model \swfour, centered on the planet.
    Top and middle rows: annihilating rates.
    Bottom row: populating rates.
    \REV{
    The net rate can be found as the \REV{lower right} plot in Figure \ref{fig:2D-SWs-Net-triplet-rates}.}
    }
    \label{fig:2D-SW4-All-triplet-rates}
\end{figure*}

Figures \ref{fig:2D-SW2-All-triplet-rates}, \ref{fig:2D-SW3-All-triplet-rates} and \ref{fig:2D-SW4-All-triplet-rates} display all the rates present in the triplet schematics for the cases \swtwo, \swthree, and \swfour\ respectively.

\begin{figure*}
    \centering
        \includegraphics[width=\linewidth]{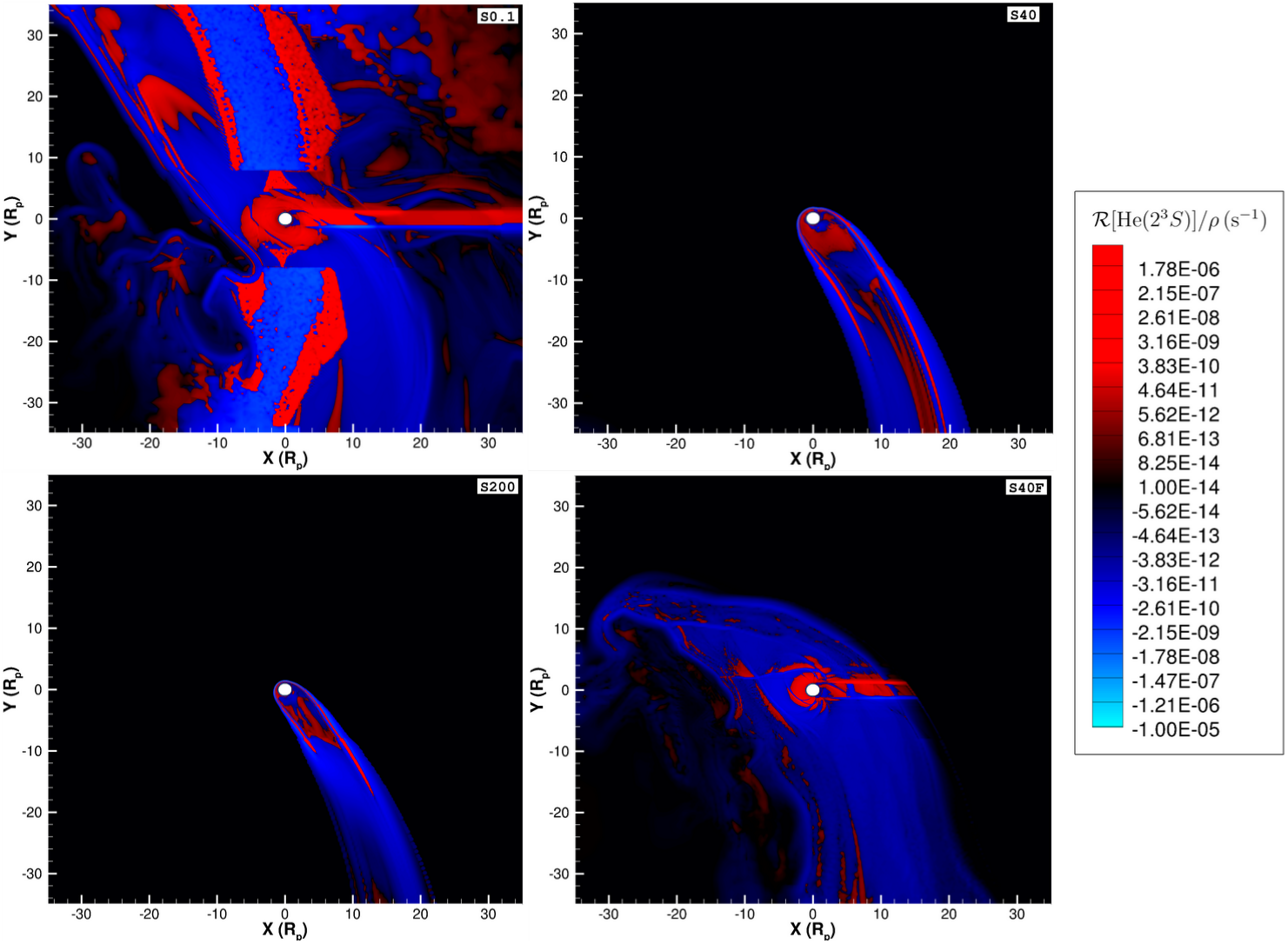}
    \REVcaption{2D view in the orbital plane $z=0$ of the net contribution from all reactions rates relative to the population of He triplet divided by the local fluid density. Models are identified at the top right corner.
    We make use of a colorbar that is logarithmic and symmetric, with black representing a null rate.}
    \label{fig:2D-SWs-Net-triplet-rates}
\end{figure*}

\begin{figure*}
    %\captionsetup{style = newfig}
    \centering
        \includegraphics[width=0.85\linewidth]{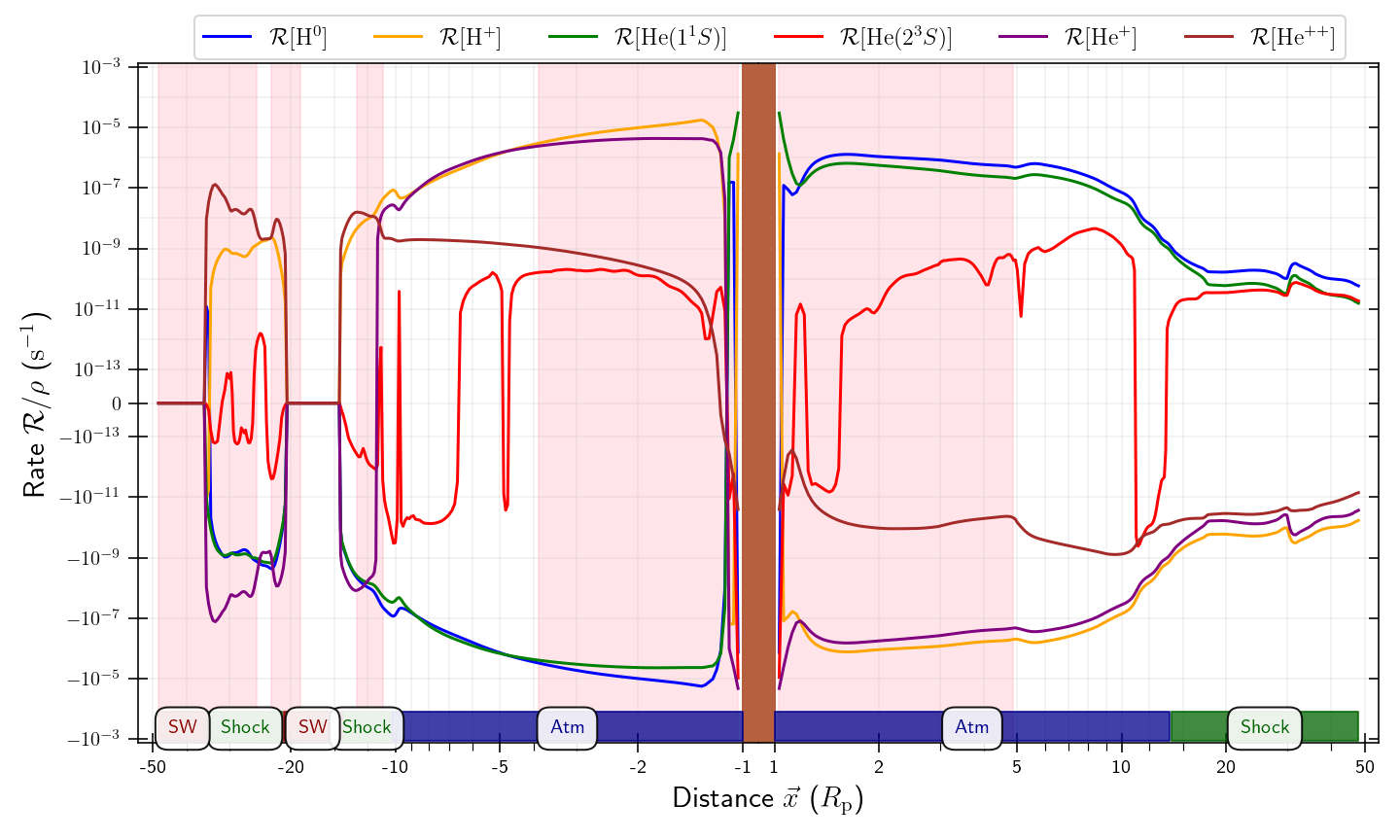}
    \REVcaption{The species net rates for the case \swone.
    As before, the y-axis shows rates in $\text{s}^{-1}$, obtained by dividing the mass volumetric rate by the total density of the fluid.
    Note that the variability in the net rate of the triplet species is imprinted in the relative fraction profile in Figure \ref{fig:2D-SWs-Net-triplet-rates}.}
    %\newfigurecaption{The species net rates for the case \swone.
    %As before, the y-axis shows rates in $\text{s}^{-1}$, obtained by dividing the mass volumetric rate by the total density of the fluid.
    %Note that the variability in the net rate of the triplet species is imprinted in the relative fraction profile in Figure \ref{fig:2D-SWs-Net-triplet-rates}.}
    \label{fig:1D-net-rates}
\end{figure*}

Figure \ref{fig:2D-SWs-Net-triplet-rates} shows the \REV{normalized} net rate for the triplet in the $xy$ plane for models \swone, \swtwo, \swthree\, and \swfour, and Figure \ref{fig:1D-net-rates} provides the entire population net-rate along the star-planet direction for model \swone, where the triplet net rate is seen as a red line.
From these figures, we can explore two distinct features: an overall smooth net-rate profile for most species, and some granulated fluctuations in $\mathcal{R}[\text{He}(2^3S)]$, \REV{mostly} pronounced in \swone. 
Overall, the triplet net rate is \REV{positive close to the planet and mostly} negative inside the expanding atmosphere. %, meaning that triplet is being annihilated.
%Nonetheless, in case \swone, we can see smooth regions of positive net rate.
For example, at \REV{ $r > 7.5\, R_p$}, there takes place a switch between the leading sink rate for triplet $\xi[\text{He}(2^1S)]$ to $\Phi [\text{He}(2^3S)]$. %, i.e., allowing for a local subdominant contribution of the sinking rates.
The region of flow-flow interaction for \swone, \swtwo\ and \swthree\ also shows net positive rates (i.e., local increase of the fraction of triplet there).
%This is a consequence of the hot stellar $\text{He}^{++}$ forcing all the remaining neutral planetary $\text{He}(1^1S)$ to become ionized.
The hot temperatures and ionized environment of the stellar wind enhances the collisional excitation rate of singlet to triplet at these boundaries.
Although models \swtwo\ and \swthree\ share similar bulk geometries, notice the growth in negative net rate \REV{volume} for the material close to the planet for the strongest stellar wind model \swthree.
\REV{We address the grid sensitivity in Section \ref{sec:Discussion-FutureImprovement}, although we emphasize that the absolute value of the net rate, $|\mathcal{R}[\text{He}(2^3S)]|$, remains smooth, with only appreciable discontinuities at shocks and grid resolution change.
}
%and they are caused by the  small number density of $\text{He}(2^3S)$ combined with a broad network of reactions and strong advection.
%The density profile of $\text{He}(2^3S)$ is the smallest for most of the grid and evolves through mass volumetric rates of the same order as the rates that originate $\text{He}^+$ or $\text{He}(1^1S)$, present at much higher densities.

%Figure \ref{fig:2D-SWs-Net-triplet-rates} shows the net value for the triplet species rate for models \swone, \swtwo, \swthree\, and \swfour. 

%Figure \ref{fig:1D-net-rates} shows the net value for the rate of each species along the $x$-axis for \swone. 

\section{Testing the importance of advection}
\label{Section_Advectionless}

Our methodology to compute the triplet population is self-consistent, that is, the triplet density is computed at every iteration and is coupled to the density, temperature, optical depth, and velocity derived in the models. 
However, this approach can become time consuming and computationally expensive. 
One solution to reduce the computation costs is to compute instead the triplet density in a post-processing fashion, i.e., the 3D models still solve for the outflow of the planet (velocity, density, temperature), but the helium population balance is not computed together with the hydrodynamical model, and instead is found \textit{a posteriori}. 
This is for example, what has been studied in the works of \citet{22_MacLeod}, which in addition, also neglects the advection term. As we showed in our study, advection plays an important role in the triplet population, for example, in the nightside, as it brings $\text{He}^+$ from elsewhere to this region.
%\REV{Figure \ref{fig:1D-net-rates} implies that advection plays an important role in setting a multi-species population.
% For example, the nightside experiences quite distinct local net rates compared to the dayside.
% The triplet's enhancement in this region (see again the first and last plot of the first row of Figure \ref{fig:2D-He-species}) can be explained by the combined transport of $\text{He}^+$ brought from elsewhere to this region and no photoionization taking place.
% This allows for a net rate that dominates over the advection rate for triplet, yet, not seen for any other species.
% }
Here, we study the effects of neglecting the advection terms and of computing the triplet population fraction ($f_{\text{He}(3)} = n_{\text{He}(2^3S)}/n_{\text{He}}$) in a post-processing step.
For that, we present two methods:
\begin{itemize}
    \item Method 1: finds $f_{\text{He}(3)}$ post-processed without assuming advection.
    \item Method 2: finds $f_{\text{He}(3)}$ post-processed without assuming advection and assuming a constant Temperature throughout the grid.
\end{itemize}

\begin{figure*}
    \centering
        \includegraphics[width=0.8\linewidth]{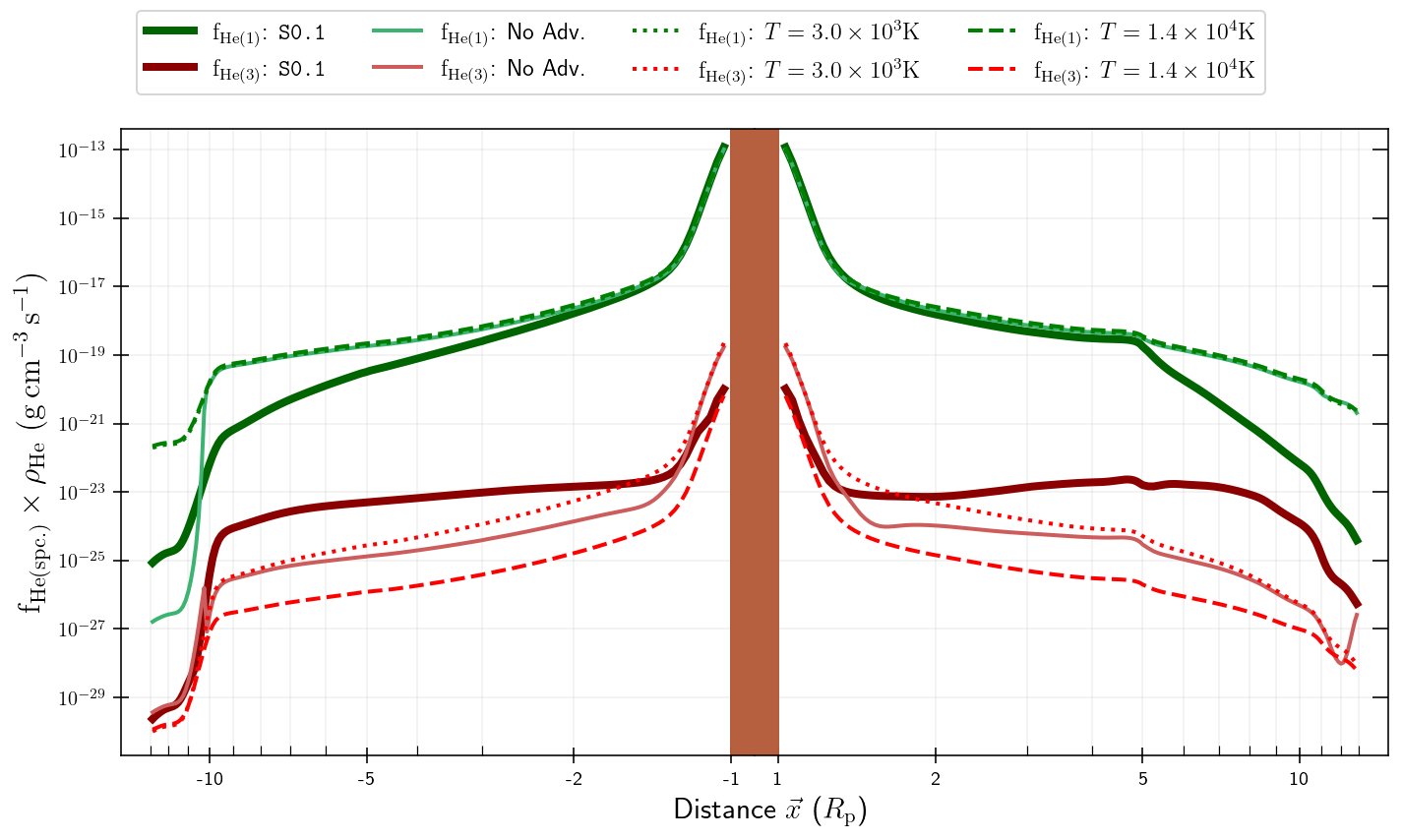}
    \REVcaption{Comparison between density of He singlet and triplet ($f_{\text{He}(1)}$, $f_{\text{He}(3)}$ respectively) from the outputs of model \swone\ (thick solid line) and post-processing methods without advection (Method 1, thin solid line) and without advection and constant temperature (Method 2, dashed and dotted lines).
    The dashed and dotted lines follow a scheme with constant temperature of $T=1.4\times10^4 \text{K}$ and $T=3.0\times10^3 \text{K}$, respectively.}
    \label{fig:Advection_methods}
\end{figure*}

We show the result of this investigation in Figure \ref{fig:Advection_methods}, where we compare our self-consistent singlet and triplet density profiles from model \swone\ and the ones derived using the alternative post-processing methods. 
Overall, neglecting advection (regardless of the method) leads to an over-population of singlet helium $f_{\text{He}(1)} = n_{\text{He}(1^1S)}/n_{\text{He}} \approx 1.0$. For the triplet population, we found that closer to the planet ($\lesssim 2\, R_p$), the triplet is over-estimated in models neglecting advection, and under-estimated beyond $\gtrsim 2\, R_p$.
%So, models that neglect advection might be under-predicting the helium transit depth. 

%%%%%%%%%%%%%%%%%%%%%%%%%%%%%%%%%%%%%%%%%%%%%%%%%%

% Don't change these lines
\bsp	% typesetting comment
\label{lastpage}
\end{document}